\documentclass[aps,prl,twocolumn,superscriptaddress,amsmath,amssymb,nofootinbib]{revtex4-1}
\usepackage{graphicx}  % needed for figures
\usepackage{dcolumn}   % needed for some tables
\usepackage{bm}        % for math
\usepackage{verbatim}   % for math
\usepackage{gensymb}
\usepackage{epstopdf}
\usepackage{footmisc}
\usepackage{natbib}
\usepackage{color}

\newcommand{\rucl}{$\alpha$-RuCl$_3$\,}
\newcommand{\Cmag}{C_\text{mag}}
\newcommand{\Smag}{S_\text{mag}}
\newcommand{\Hc}{H_\text{c}}
\newcommand{\Hct}{H_{\text{c}2}}
\newcommand{\JKG}{$J_1$--$K_1$--$\Gamma_1$--$J_3$}

\begin{document}

\title{Field-induced quantum criticality in the Kitaev system $\alpha$-RuCl$_3$}

\author{A. U. B. Wolter}
\author{L. T. Corredor*}
\affiliation{Leibniz-Institut f\"ur Festk\"orper- und
Werkstoffforschung (IFW) Dresden, 01171 Dresden, Germany}
\thanks{These authors contributed equally to this work.}
\author{L. Janssen*}
\affiliation{Institut f\"ur Theoretische Physik, Technische
Universit\"at Dresden, 01062 Dresden, Germany}
\author{K. Nenkov}
\affiliation{Leibniz-Institut f\"ur Festk\"orper- und
Werkstoffforschung (IFW) Dresden, 01171 Dresden, Germany}
\author{S. Sch\"{o}necker}
\affiliation{Department of Materials Science and Engineering, KTH
- Royal Institute of Technology, Stockholm 10044, Sweden}
\author{S.-H. Do}
\author{K.-Y. Choi}
\affiliation{Department of Physics, Chung-Ang University, Seoul
156-756, Republic of Korea}
\author{R. Albrecht}
\author{J. Hunger}
\author{T. Doert}
\affiliation{Fachrichtung Chemie und Lebensmittelchemie,
Technische Universit\"at Dresden, 01062 Dresden, Germany}
\author{M. Vojta}
\affiliation{Institut f\"ur Theoretische Physik, Technische
Universit\"at Dresden, 01062 Dresden,  Germany}
\author{B. B\"{u}chner}
\affiliation{Leibniz-Institut f\"ur Festk\"orper- und
Werkstoffforschung (IFW) Dresden, 01171 Dresden, Germany}
\affiliation{Institut f\"ur Festk\"orperphysik, Technische
Universit\"{a}t Dresden, 01062 Dresden, Germany}

\date{\today}

%%%%%%%%%%%%%%%%%%%%%%%%%%%%%%%%%%%%%%%%%%%%%%%%%%%%%%%%%%%%%%%%%%%%%%%%%%%%%%%%%%%

\begin{abstract}
\rucl has attracted enormous attention since it has been proposed
as a prime candidate to study fractionalized magnetic excitations
akin to Kitaev's honeycomb-lattice spin liquid. We have performed
a detailed specific-heat investigation at temperatures down to
$0.4$~K in applied magnetic fields up to $9$~T for fields parallel
to the $ab$ plane. We find a suppression of the zero-field
antiferromagnetic order, together with an increase of the
low-temperature specific heat, with increasing field up to
$\mu_0\Hc\approx 6.9$~T. Above $\Hc$, the magnetic contribution to
the low-temperature specific heat is strongly suppressed, implying
the opening of a spin-excitation gap. Our data point toward a
field-induced quantum critical point (QCP) at $\Hc$; this is
supported by universal scaling behavior near $\Hc$. Remarkably,
the data also reveal the existence of a small characteristic
energy scale well below $1$~meV above which the excitation
spectrum changes qualitatively. We relate the data to theoretical
calculations based on a {\JKG} honeycomb model.
\end{abstract}

\maketitle

%%%%%%%%%%%%%%%%%%%%%%%%%%%%%%%%%%%%%%%%%%%%%%%%%%%%%%%%%%%%%%%%%%%%%%%%%%%%%%%%%%%
%--------Introduction

\rucl is a $J_{\rm eff}\!=\!1/2$ Mott insulator with a layered
structure of edge-sharing RuCl$_6$ octahedra arranged in a
honeycomb lattice
\cite{Plumb2014,Sears2015,Johnson2015,Majumder2015,Kubota2015,Sinn2016,Ziatdinov2016,Weber2016}.
It has been suggested \cite{Jackeli2009,Chaloupka2010} that
strongly spin-orbit-coupled Mott insulators with that lattice
geometry realize bond-dependent magnetic ``compass'' interactions
\cite{Nussinov_RMP} which, if dominant, would lead to a quantum
spin liquid (QSL) ground state as discussed by Kitaev
\cite{Kitaev2006}. This exotic spin-disordered state displays an
emergent $Z_2$ gauge field and fractionalized Majorana-fermion
excitations relevant for topological quantum computation
\cite{Kitaev2006,Sandilands2015,Nasu2015,Trebst2017}.

While \rucl displays magnetic long-range order (LRO) of so-called
zigzag type, it has been proposed to be proximate to the Kitaev
spin liquid based on its small ordering temperature and its
unusual magnetic excitation spectrum
\cite{Singh2012,Banerjee2016,Cao2016}. The magnetic interactions
between the Ru$^{3+}$ magnetic moments are believed to be
described by a variant of the Heisenberg-Kitaev model
\cite{Chaloupka2010}: Electronic-structure calculations indicate
that the Kitaev interaction in \rucl is ferromagnetic and indeed
defines the largest exchange energy scale
\cite{Yadav2016,Winter2016}. However, the debate about the spin
model most appropriate for \rucl -- likely to include Heisenberg
and off-diagonal exchange interactions, possibly also beyond
nearest neighbors -- has not been settled
\cite{Kimchi2011,Rau2014,Yamaji2014,Reuther2014,Perkins2014,Rousso2015,Yadav2016,Winter2016,Winter2017,Laubach2017}.

The physics of \rucl in an external magnetic field promises to be
particulary interesting: It has been reported
\cite{Leahy2016,Baek2017} that magnetic ordering disappears for
fields of the order of $10$~T (depending on the field direction),
and NMR measurements performed down to $4$~K have indicated the
formation of a sizeable spin gap at high fields \cite{Baek2017}.
Additionally, numerical exact-diagonalization studies of an
extended Heisenberg-Kitaev model found hints for a transition from
zigzag magnetic ordering to a spin-liquid state when applying a
magnetic field \cite{Yadav2016}.

In this Letter, we report a careful heat-capacity study of \rucl
down to low temperature $T$ of $0.4$~K in in-plane fields up to
$9$~T. We confirm the field-induced suppression of LRO at a
critical field of $\mu_0\Hc\approx 6.9$~T and provide a detailed
account of the field evolution of the spin gap: This is small
below $H_c$, closes at $H_c$, and progressively grows above $\Hc$.
The specific-heat data displays universal scaling consistent with
the existence of a quantum critical point (QCP) at $\Hc$. The
scaling analysis yields critical exponents $d/z = 2.1\pm0.1$ and
$\nu z = 0.7\pm0.1$ where $d$ is the space dimension and $\nu$ and
$z$ are the correlation-length and dynamic critical exponents,
respectively.
Based on explicit calculations for a {\JKG} spin model we argue
that the specific-heat behavior near $\Hc$ implies a mode
softening at $\Hc$ that accompanies the disappearance of magnetic
order.
The observed violations of scaling for $T \ge 3$~K indicate the
presence of an intrinsic sub-meV energy scale near the QCP which
we interpret as signature of Kitaev physics.

%%%%%%%%%%%%%%%%%%%%%%%%%%%%%%%%%%%%%%%%%%%%%%%%%%%%%%%%%%%%%%%%%%%%%%%%%%%%%%%%%%%
%-----------Experimental details

\paragraph{Experimental:}
High-quality single crystals of \rucl were grown by a vacuum
sublimation method. A commercial RuCl$_3$ powder (Alfa-Aesar) was
thoroughly grounded, and dehydrated in a quartz ampoule at
$250$\degree C for two days. The ampoule was sealed in vacuum and
placed in a temperature-gradient furnace. The temperature of the
RuCl$_3$ powder was set at $1080$\degree C. After five hours the
furnace was cooled to $600$\degree C at a rate of $-2$\degree C/h.
The magnetic properties of the crystal were checked through
measurements as a function of $T$ and $H$ using a Vibrating Sample
Magnetometer (Quantum Design) with SQUID detection (SQUID-VSM),
see supplement \cite{suppl} for the magnetic characterization.
Specific-heat measurements were performed on a single crystal ($m
\sim 7$ mg) between $0.4$ K and $20$~K using a heat-pulse
relaxation method in a Physical Properties Measurement System
(PPMS, Quantum Design), in magnetic fields up to $9$~T parallel to
the $ab$ plane.

%%%%%%%%%%%%%%%%%%%%%%%%%%%%%%%%%%%%%%%%%%%%%%%%%%%%%%%%%%%%%%%%%%%%%%%%%%%%%%%%%%%
%----------Results and discussion

\begin{figure}
\includegraphics[scale=0.5]{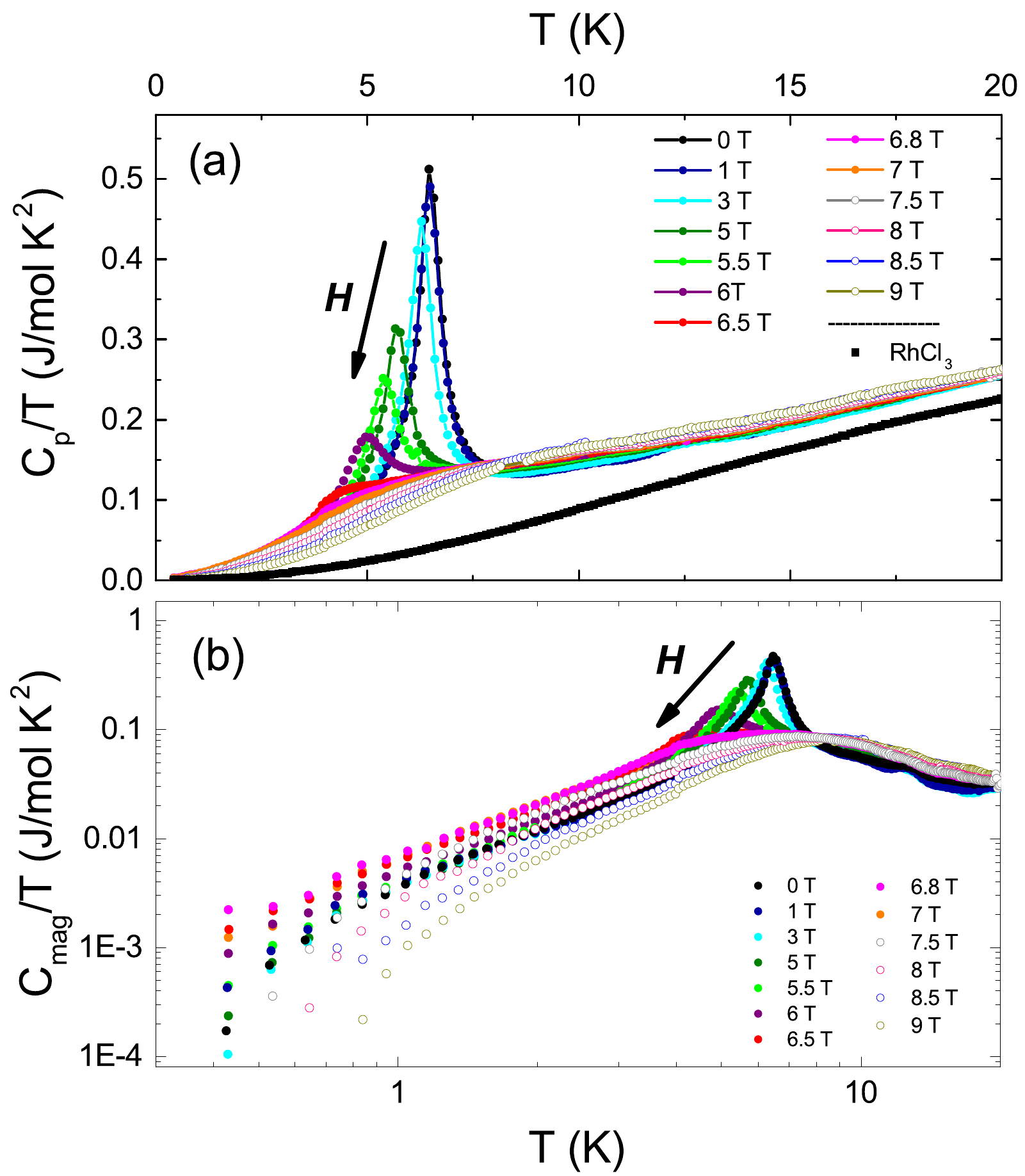}
\caption{\label{Cp}(color online) (a) Temperature dependence of
the specific heat, plotted as $C_p/T$, of \rucl for different
magnetic fields up to 9 T $\parallel ab$.
(b) As before, but showing the magnetic contribution to the
specific heat after phonon subtraction on a log-log scale, for
details see text. }
\end{figure}

\paragraph{Results:}
The low-$T$ specific heat $C_p/T$ as a function of temperature in
different applied fields is shown in Fig. \ref{Cp}(a). The
zero-field curve reveals the good quality of the sample, with a
single magnetic transition at $T_N = 6.5$~K determined from the
peak position. By applying a magnetic field the peak becomes
broader and the transition temperature is gradually suppressed.
Finally no thermal phase transition is detected for fields higher
than $6.9$~T, i.e., magnetic LRO disappears.

In order to extract the magnetic contribution to the low-$T$
specific heat, the data were analyzed by subtracting the lattice
contribution from the experimental $C_p(T)$ data by measuring the
non-magnetic structural analog compound RhCl$_3$ in pressed
polycrystalline form. The difference of mass and volume between
the Rh and Ru compounds was accounted for by scaling the
experimental specific heat curve by the Lindemann factor
\cite{Lindemann1910}, which was found to be 0.98.
With the aim of ruling out possible errors due to non-perfect
sample coupling during the measurements, the phononic contribution
was also calculated for RhCl$_3$ by density-functional theory, see
supplement ~\cite{suppl}. This approach confirmed that the phonon
subtraction based on the experimental data is consistent with the
theoretical calculations for $T \ge 1$~K.

The temperature dependence of the calculated magnetic contribution
to the specific heat is shown in Fig. \ref{Cp}(b). In the
lowest-$T$ region, $T \leq 3$~K, an increase of $\Cmag/T$ with the
applied field could be observed up to $\mu_0 H = 6.8$ T.
Increasing the field even further, the opposite behavior is
revealed: the magnetic contribution starts to decrease with field
up to the highest field of $9$~T. Hence, low-$T$ entropy
accumulates around $6.8-7$~T.
Remarkably, around $6.9$~T the magnetic specific heat displays an
approximate power-law behavior between $0.4$ and $2.5$~K, with
$\Cmag \propto T^x$ with $x\approx2.5$. Together, these
observations imply the existence of a field-induced QCP
\cite{ssbook,Vojta2003} at $\mu_0\Hc\approx 6.9$~T.

%%%%%%%%%%%%%%%%%%%%%%%%%%%%%%%%%%%%%%%%%%%%%%%%%%%%%%%%%%%%%%%%%%%%%%%%%%%%%%%%%%%

\begin{figure}
\includegraphics[scale=0.45]{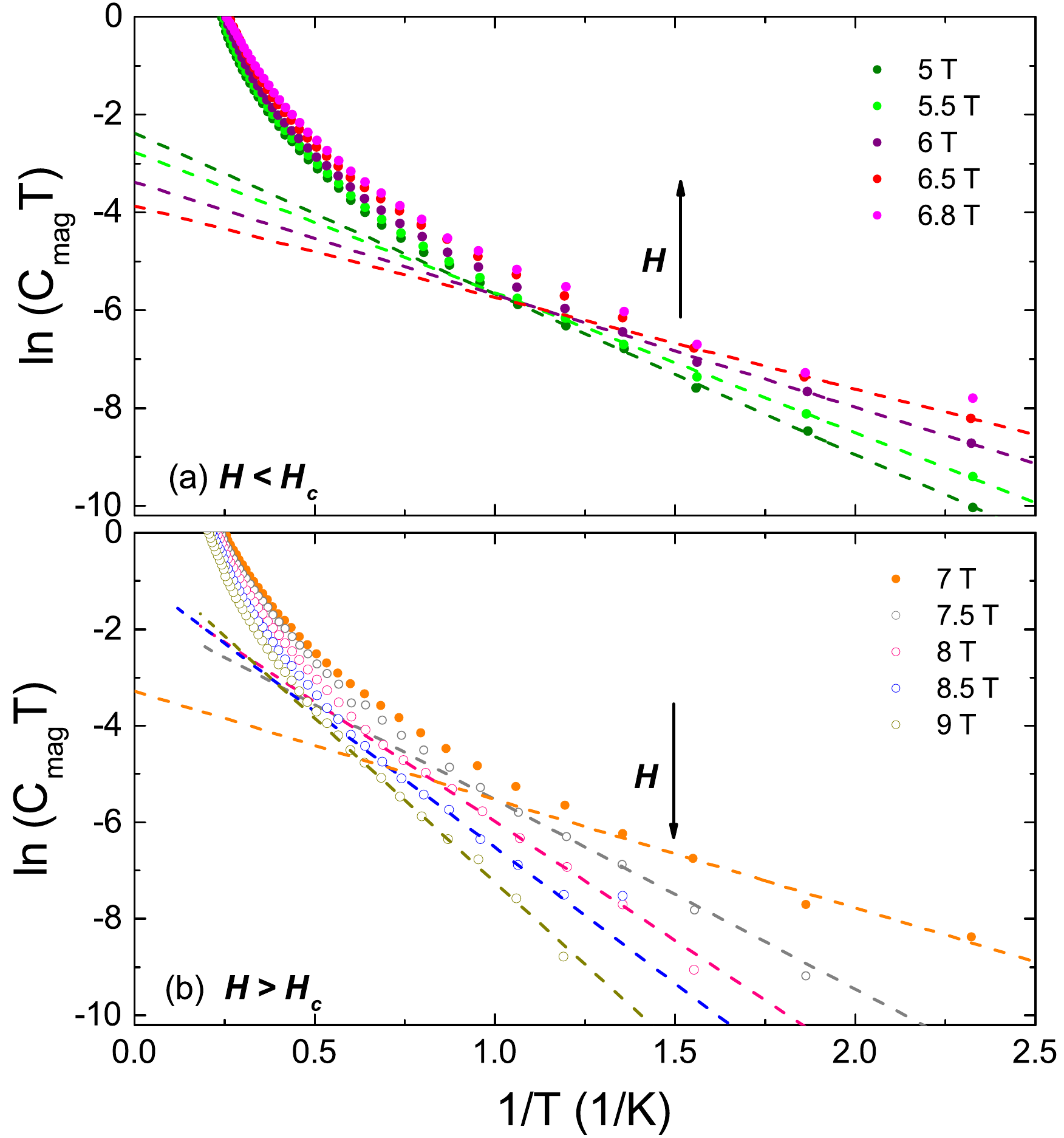}
\caption{\label{Cmagfits}(color online) Exponential fit of $\Cmag
T$ in order to extract the excitation gap for magnetic fields (a)
$5$~T $\leq \mu_0 H \leq 6.8$~T and (b) $7$~T $\leq \mu_0H \leq
9$~T. The data at $6.8$~T cannot be meaningfully fit by an
exponential, i.e., the gap is too small. }
\end{figure}

\paragraph{Excitation gap:}
The lowest-temperature data away from the QCP, with a gradual
suppression of $\Cmag(T)$, indicate the opening of a magnetic
excitation gap, Fig. \ref{Cp}(b). The simplest model of a bosonic
mode with gap $\Delta$ and parabolic dispersion in $d=2$ predicts
that $\Cmag \propto \exp[-\Delta/(k_BT)]/T$, see supplement
\cite{suppl}. According to this, the experimental $\Cmag T$ data
were fitted to a pure exponential behavior in order to extract the
energy gap. The results are shown in Fig. \ref{Cmagfits}.

Two key observations are apparent:
First, the data below about $1.5$~K indeed show an exponential
suppression of $\Cmag$, and the corresponding gap is minimal near
the putative QCP at $\mu_0\Hc\approx 6.9$~T. It varies
monotonically on both sides of the QCP, consistent with
theoretical expectations \cite{ssbook,Vojta2003}. (Note that a
symmetry-broken phase below $\Hc$ should also display a gap, as no
Goldstone modes are expected due to the presence of strong
spin-orbit coupling.)
Second, the data above $\sim$ $1.5$~K do {\em not} follow an
exponential behavior (at least not in the field range studied
here); in fact $\Cmag/T$ between $1.5$ and $5$~K appears more
consistent with a power law, Fig. \ref{Cp}(b). This indicates that
the density of states of magnetic excitations changes its
character at a small energy scale of a few tenths of a meV
\cite{suppl}.

%%%%%%%%%%%%%%%%%%%%%%%%%%%%%%%%%%%%%%%%%%%%%%%%%%%%%%%%%%%%%%%%%%%%%%%%%%%%%%%%%%%

\begin{figure}
\includegraphics[scale=0.48]{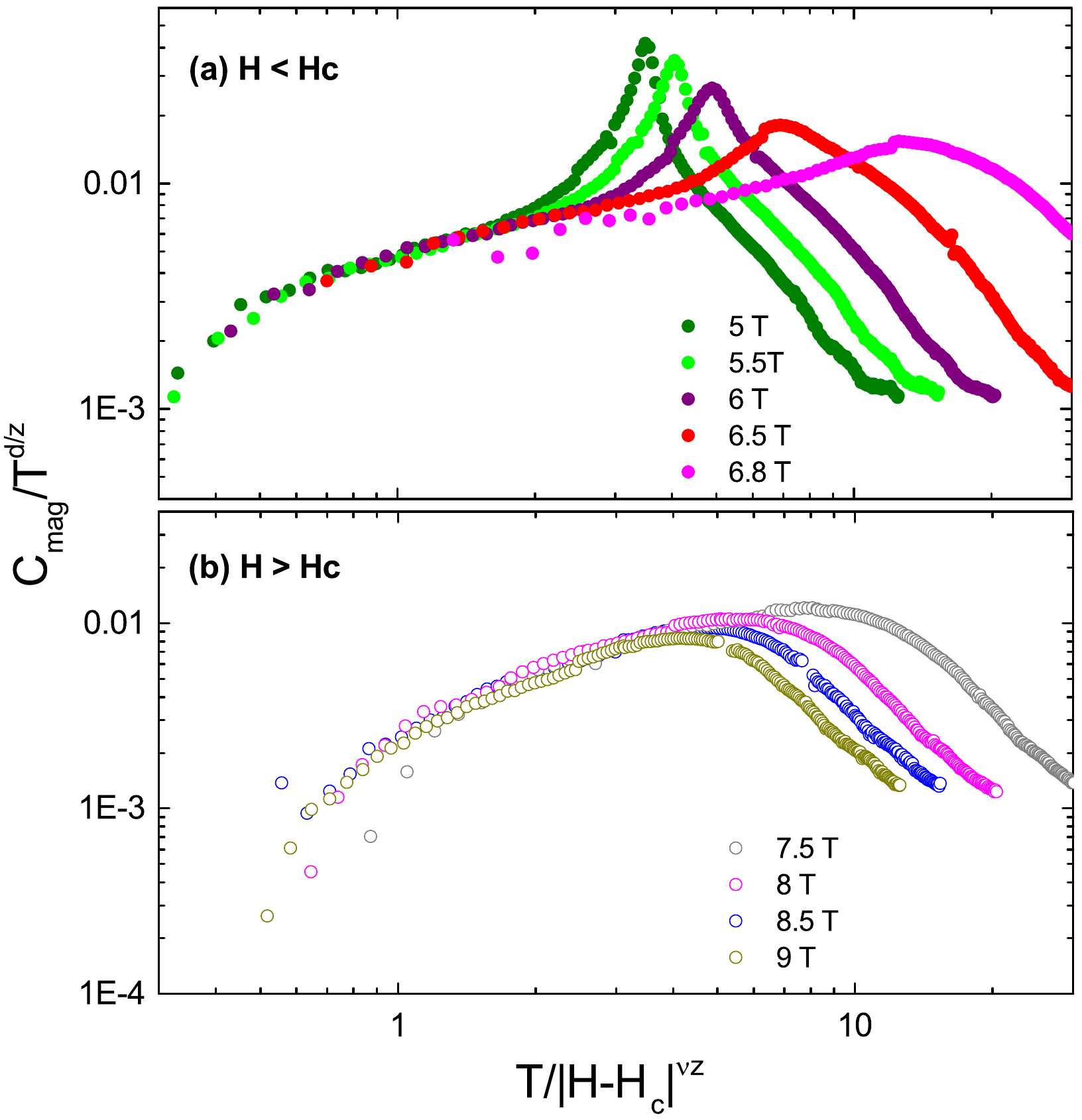}
\caption{\label{scaling}(color online) Scaling plot of
$\Cmag(T,H)$, showing $\Cmag/T^{d/z}$ versus $T/(H - \Hc)^{\nu
z}$, here for $\mu_0\Hc=7$~T, $d/z=2.1$, and $\nu z=0.7$. The two
panels shows fields (a) slightly below and (b) slightly above
$\Hc$. The universal piece in the upper (lower) panel corresponds
to the scaling function $f_-$ ($f_+$) in Eq.~\eqref{eq:scale}.}
\end{figure}

\paragraph{Scaling analysis:}
In order to further substantiate the QCP hypothesis, we have
performed a scaling analysis of $\Cmag(T,H)$. Provided that
hyperscaling holds, the critical contribution to the specific heat
is expected \cite{ssbook,Vojta2003} to scale as
\begin{equation}
\label{eq:scale} C = T^{d/z} f_\pm(T/|H-\Hc|^{\nu z}),
\end{equation}
where $f_\pm$ are universal functions describing the scaling for
$H>\Hc$ and $H<\Hc$, respectively, and the argument
$T/|H-\Hc|^{\nu z}$ is made dimensionless by using suitable units.
Plotting the specific heat as $C/T^{d/z}$ as a function of
$T/|H-\Hc|^{\nu z}$, separately for $H\lesssim \Hc$ and $H\gtrsim
\Hc$, we find an approximate data collapse for $d/z = 2.1\pm0.1$,
$\nu z = 0.7\pm0.1$, and $\mu_0H_c=6.9\pm0.1$~T, see
Fig.~\ref{scaling} for an example. (Note that the data cannot be
collapsed with $d/z=2.5$.) For comparison, the supplemental
Fig.~S5(c) shows the scaling collapse of specific-heat data
obtained from a spin-wave-based model calculation for a
field-driven QCP in a {\JKG} model, for details see
Ref.~\onlinecite{suppl}. The agreement reinforces the notion of a
field-induced QCP in \rucl.

It is instructive to analyze deviations from scaling in
Fig.~\ref{scaling}:
(i) None of the data sets realizes the critical power law $\Cmag
\propto T^{d/z}$, indicating that the critical point has not been
reached precisely. The most likely reason is sample
inhomogeneities, e.g., caused by crystallographic domains with
different in-plane orientation.  These would lead to a
distribution of $|H-\Hc|$ values due to anisotropic $g$ factors
and hence to a smearing of the QCP.
(ii) Only data below $3$~K follow the approximate scaling; this is
particularly clear from Fig.~\ref{scaling}(a) where the
specific-heat peaks corresponding to $T_N$ do not scale. This
again implies the existence of a small energy scale, only below
which standard quantum critical scaling applies.

%%%%%%%%%%%%%%%%%%%%%%%%%%%%%%%%%%%%%%%%%%%%%%%%%%%%%%%%%%%%%%%%%%%%%%%%%%%%%%%%%%%

\begin{figure}
\includegraphics[scale=0.33]{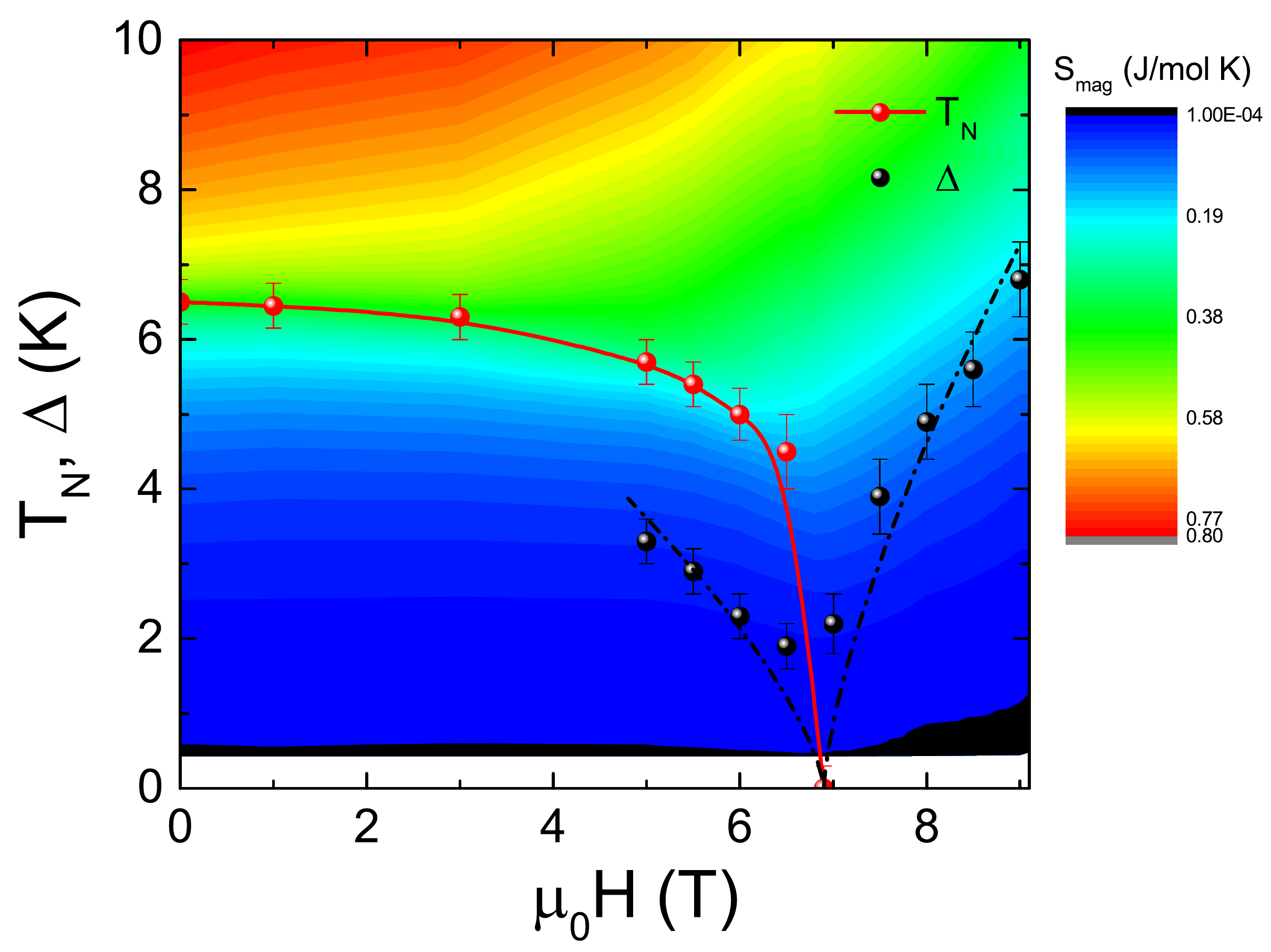}
\caption{\label{PD}(color online) $T$-$H$ phase diagram for \rucl:
magnetic ordering temperature and energy gap as a function of the
applied magnetic field $\parallel ab$. The dashed line corresponds
to the fit of the gap function to $\Delta \propto |H -
\Hc|^{0.7}$. Additionally, the magnetic entropy $\Smag(T,H)$ is
shown in color scale.}
\end{figure}

\paragraph{Phase diagram:}
Our findings are summarized in the phase diagram, Fig.~\ref{PD},
which displays the N\'eel temperature (from the peak position in
$C_p/T$ as a function of $T$) and the gap values extracted as in
Fig.~\ref{Cmagfits}. The loss of magnetic order at $\Hc$ is
accompanied by the closing of the magnetic excitation gap
$\Delta$. Figure~\ref{PD} also shows the magnetic entropy $\Smag$,
obtained from integrating the specific-heat data from Fig.
\ref{Cp}(b). Focussing on $\Smag(H)$ at fixed $T$, the entropy
accumulation near $\Hc$ is clearly visible, as is the gap
formation at elevated fields.

According to standard scaling, the gap values should follow a
power law $\Delta \propto |H - \Hc|^{\nu z}$. This is
approximately obeyed by the experimental data with $\nu z=0.7$,
but deviations are visible very close to $\Hc$.
These deviations could in principle arise from the transition
being weakly first order (in which case the gap would not vanish
at $\Hc$). We have checked this possibility by performing field
sweeps at $1.8$~K searching for hysteresis \cite{suppl}. However,
the detected hysteresis in $M(H)$ is tiny, presumably arising from
defects, such that we can exclude intrinsic first-order behavior.
Hence, the deviations from power laws likely originate from sample
inhomogeneities as discussed above. Alternatively, the formation
of an additional narrow low-$T$ phase near $\Hc$ appears possible,
as theoretically predicted in Ref.~\onlinecite{Janssen2016} for
the classical Heisenberg-Kitaev model; this requires more detailed
low-$T$ measurements as a function of continuous $H$.

%%%%%%%%%%%%%%%%%%%%%%%%%%%%%%%%%%%%%%%%%%%%%%%%%%%%%%%%%%%%%%%%%%%%%%%%%%%%%%%%%%%

\paragraph{Mode softening and nature of the high-field phase:}
We now return to the specific-heat data and discuss them in the
context of theoretical scenarios for the quantum phase transition
(QPT) at $\Hc$.
The data show that LRO is lost above $\Hc$. If the QPT at $\Hc$ is
continuous then this should be accompanied with a soft mode, i.e.,
the high-field phase should display a gapped mode with gap
$\Delta\to0$ as $H\to \Hc^+$, with this mode condensation
establishing zigzag LRO below $\Hc$. The specific-heat data above
$\Hc$ is consistent with these considerations.

An exciting possibility is that the phase above $\Hc$ is a
field-induced spin liquid, accompanied by topological order. Then,
the mode which softens at $\Hc$ would presumably correspond to an
excitation of the emergent gauge field (dubbed vison for a $Z_2$
spin liquid). The field-induced spin liquid cannot exist up to
arbitrarily high fields, i.e., a second QPT at a higher field
$\Hct$ should exist where the spin liquid is destroyed in favor of
the high-field phase; this has not been experimentally tested to
date. While indications for a field-induced spin liquid in
Heisenberg-Kitaev models were found in numerical simulations in
Ref.~\onlinecite{Yadav2016}, a full theory is not available.

Alternatively, the phase above $\Hc$ could be adiabatically
connected to the high-field limit, and the soft mode would then
correspond to a high-field magnon. We note that such a magnon
condensation is rather different from that in an SU(2)-symmetric
Heisenberg magnet due to spin-orbit coupling: First, the
zero-temperature magnetisation above $\Hc$ can be far below
saturation. Second, due to the low symmetry the QPT is not of BEC
type ($z=2$, $\nu=1/2$) but generically in the Ising universality
class ($z=1$, $\nu=0.630$ in $d=2$) .

We have studied this type of magnon-condensation transition in the
framework of an appropriate {\JKG} model \cite{Winter2016} in some
detail, see supplement \cite{suppl}. Within our semiclassical
approach, the critical exponents of the transition are $\nu=1/2$
and $z=1$. The results \cite{suppl}, including the value of $\Hc$,
appear in semiquantitative agreement with the experimental data.
This lends further credit to the presence of a field-induced QCP
in \rucl, but does not allow us to conclusively identify the
nature of the high-field phase.
We also note that the theoretical calculation shows the presence
of an additional energy scale arising from strong van-Hove
singularities in the magnon band structure at high fields. This
energy scale varies approximately linearly with field above $\Hc$
but does not vanish at $\Hc$, see Fig.~S6. Beyond the
semiclassical limit these elevated-energy features are likely to
loose their sharp-mode character, possibly due to
fractionalization, as has been found in related models at zero
field \cite{Gohlke2017}.

%%%%%%%%%%%%%%%%%%%%%%%%%%%%%%%%%%%%%%%%%%%%%%%%%%%%%%%%%%%%%%%%%%%%%%%%%%%%%%%%%%%

\paragraph{Summary:}
Via low-temperature specific heat measurements we have
demonstrated that the frustrated magnet \rucl displays
field-induced quantum criticality at $\mu_0\Hc \approx 6.9$~T
applied in the $ab$ plane. The high-field phase is characterized
by a field-induced gap to magnetic excitations which is clearly
visible below $\sim 2$~K. Our scaling analysis of the low-$T$
specific-heat data yields estimates for the critical exponents
$d/z = 2.1\pm0.1$ and $\nu z = 0.7\pm0.1$, consistent with Ising
universality. While we cannot draw conclusions about the nature of
the high-field phase, we believe that the hypothesis of a
field-induced spin liquid deserves further studies.

Importantly, the data also reveal the existence of a sub-meV
energy scale near the QCP above which the nature of the excitation
spectrum changes. It is conceivable that this scale corresponds to
a crossover from more conventional dispersive modes at low
energies to exotic fractionalized excitations driven by Kitaev
interactions. Studying the evolution of these excitations at
higher fields is an exciting task for the future.

%%%%%%%%%%%%%%%%%%%%%%%%%%%%%%%%%%%%%%%%%%%%%%%%%%%%%%%%%%%%%%%%%%%%%%%%%%%%%%%%%%%

We acknowledge insightful discussions with C. Hess, A. Isaeva, R.
Moessner, S. Nagler, F. Pollmann, S. Rachel, M. Richter, and J.
van den Brink.
The phonon simulations were performed on resources provided by the
Swedish National Infrastructure for Computing (SNIC) at the
supercomputer centers in Link\"oping and Stockholm.
This research has been supported by the DFG via SFB 1143.

%%%%%%%%%%%%%%%%%%%%%%%%%%%%%%%%%%%%%%%%%%%%%%%%%%%%%%%%%%%%%%%%%%%%%%%%%%%%%%%%%%%

\paragraph{Note added:}
While this paper was being written, parallel work
\cite{Sears2017,Zheng2017} appeared on arXiv documenting related
studies of \rucl in a magnetic field. While
Ref.~\onlinecite{Sears2017} reported gapped magnetic excitations
at fields above $\Hc$, the results of Ref.~\onlinecite{Zheng2017}
were interpreted in terms of gapless excitations in this regime.
Interestingly, Ref.~\onlinecite{Sears2017} quotes the
order-parameter exponent at $\Hc$ to be $\beta=0.28$, in
reasonable agreement with the Ising value $0.326$, suggesting a
conventional Ising transition. However, in both
Refs.~\onlinecite{Sears2017,Zheng2017} the measurements were
restricted to temperatures above $2$~K. Our data show that lower
temperatures are required to reach the asymptotic regime.

\bibliography{aRuCl3_Paper_v1_LT}

\begin{thebibliography}{38}%
\makeatletter
\providecommand \@ifxundefined [1]{%
 \@ifx{#1\undefined}
}%
\providecommand \@ifnum [1]{%
 \ifnum #1\expandafter \@firstoftwo
 \else \expandafter \@secondoftwo
 \fi
}%
\providecommand \@ifx [1]{%
 \ifx #1\expandafter \@firstoftwo
 \else \expandafter \@secondoftwo
 \fi
}%
\providecommand \natexlab [1]{#1}%
\providecommand \enquote  [1]{``#1''}%
\providecommand \bibnamefont  [1]{#1}%
\providecommand \bibfnamefont [1]{#1}%
\providecommand \citenamefont [1]{#1}%
\providecommand \href@noop [0]{\@secondoftwo}%
\providecommand \href [0]{\begingroup \@sanitize@url \@href}%
\providecommand \@href[1]{\@@startlink{#1}\@@href}%
\providecommand \@@href[1]{\endgroup#1\@@endlink}%
\providecommand \@sanitize@url [0]{\catcode `\\12\catcode `\$12\catcode
  `\&12\catcode `\#12\catcode `\^12\catcode `\_12\catcode `\%12\relax}%
\providecommand \@@startlink[1]{}%
\providecommand \@@endlink[0]{}%
\providecommand \url  [0]{\begingroup\@sanitize@url \@url }%
\providecommand \@url [1]{\endgroup\@href {#1}{\urlprefix }}%
\providecommand \urlprefix  [0]{URL }%
\providecommand \Eprint [0]{\href }%
\providecommand \doibase [0]{http://dx.doi.org/}%
\providecommand \selectlanguage [0]{\@gobble}%
\providecommand \bibinfo  [0]{\@secondoftwo}%
\providecommand \bibfield  [0]{\@secondoftwo}%
\providecommand \translation [1]{[#1]}%
\providecommand \BibitemOpen [0]{}%
\providecommand \bibitemStop [0]{}%
\providecommand \bibitemNoStop [0]{.\EOS\space}%
\providecommand \EOS [0]{\spacefactor3000\relax}%
\providecommand \BibitemShut  [1]{\csname bibitem#1\endcsname}%
\let\auto@bib@innerbib\@empty
%</preamble>
\bibitem [{\citenamefont {Plumb}\ \emph {et~al.}(2014)\citenamefont {Plumb},
  \citenamefont {Clancy}, \citenamefont {Sandilands}, \citenamefont {Shankar},
  \citenamefont {Hu}, \citenamefont {Burch}, \citenamefont {Kee},\ and\
  \citenamefont {Kim}}]{Plumb2014}%
  \BibitemOpen
  \bibfield  {author} {\bibinfo {author} {\bibfnamefont {K.~W.}\ \bibnamefont
  {Plumb}}, \bibinfo {author} {\bibfnamefont {J.~P.}\ \bibnamefont {Clancy}},
  \bibinfo {author} {\bibfnamefont {L.~J.}\ \bibnamefont {Sandilands}},
  \bibinfo {author} {\bibfnamefont {V.~V.}\ \bibnamefont {Shankar}}, \bibinfo
  {author} {\bibfnamefont {Y.~F.}\ \bibnamefont {Hu}}, \bibinfo {author}
  {\bibfnamefont {K.~S.}\ \bibnamefont {Burch}}, \bibinfo {author}
  {\bibfnamefont {H.-Y.}\ \bibnamefont {Kee}}, \ and\ \bibinfo {author}
  {\bibfnamefont {Y.-J.}\ \bibnamefont {Kim}},\ }\href {\doibase
  10.1103/physrevb.90.041112} {\bibfield  {journal} {\bibinfo  {journal} {Phys.
  Rev. B}\ }\textbf {\bibinfo {volume} {90}},\ \bibinfo {pages} {041112}
  (\bibinfo {year} {2014})}\BibitemShut {NoStop}%
\bibitem [{\citenamefont {Sears}\ \emph {et~al.}(2015)\citenamefont {Sears},
  \citenamefont {Songvilay}, \citenamefont {Plumb}, \citenamefont {Clancy},
  \citenamefont {Qiu}, \citenamefont {Zhao}, \citenamefont {Parshall},\ and\
  \citenamefont {Kim}}]{Sears2015}%
  \BibitemOpen
  \bibfield  {author} {\bibinfo {author} {\bibfnamefont {J.~A.}\ \bibnamefont
  {Sears}}, \bibinfo {author} {\bibfnamefont {M.}~\bibnamefont {Songvilay}},
  \bibinfo {author} {\bibfnamefont {K.~W.}\ \bibnamefont {Plumb}}, \bibinfo
  {author} {\bibfnamefont {J.~P.}\ \bibnamefont {Clancy}}, \bibinfo {author}
  {\bibfnamefont {Y.}~\bibnamefont {Qiu}}, \bibinfo {author} {\bibfnamefont
  {Y.}~\bibnamefont {Zhao}}, \bibinfo {author} {\bibfnamefont {D.}~\bibnamefont
  {Parshall}}, \ and\ \bibinfo {author} {\bibfnamefont {Y.-J.}\ \bibnamefont
  {Kim}},\ }\href {\doibase 10.1103/physrevb.91.144420} {\bibfield  {journal}
  {\bibinfo  {journal} {Phys. Rev. B}\ }\textbf {\bibinfo {volume} {91}},\
  \bibinfo {pages} {144420} (\bibinfo {year} {2015})}\BibitemShut {NoStop}%
\bibitem [{\citenamefont {Johnson}\ \emph {et~al.}(2015)\citenamefont
  {Johnson}, \citenamefont {Williams}, \citenamefont {Haghighirad},
  \citenamefont {Singleton}, \citenamefont {Zapf}, \citenamefont {Manuel},
  \citenamefont {Mazin}, \citenamefont {Li}, \citenamefont {Jeschke},
  \citenamefont {Valent{\'{\i}}},\ and\ \citenamefont {Coldea}}]{Johnson2015}%
  \BibitemOpen
  \bibfield  {author} {\bibinfo {author} {\bibfnamefont {R.~D.}\ \bibnamefont
  {Johnson}}, \bibinfo {author} {\bibfnamefont {S.~C.}\ \bibnamefont
  {Williams}}, \bibinfo {author} {\bibfnamefont {A.~A.}\ \bibnamefont
  {Haghighirad}}, \bibinfo {author} {\bibfnamefont {J.}~\bibnamefont
  {Singleton}}, \bibinfo {author} {\bibfnamefont {V.}~\bibnamefont {Zapf}},
  \bibinfo {author} {\bibfnamefont {P.}~\bibnamefont {Manuel}}, \bibinfo
  {author} {\bibfnamefont {I.~I.}\ \bibnamefont {Mazin}}, \bibinfo {author}
  {\bibfnamefont {Y.}~\bibnamefont {Li}}, \bibinfo {author} {\bibfnamefont
  {H.~O.}\ \bibnamefont {Jeschke}}, \bibinfo {author} {\bibfnamefont
  {R.}~\bibnamefont {Valent{\'{\i}}}}, \ and\ \bibinfo {author} {\bibfnamefont
  {R.}~\bibnamefont {Coldea}},\ }\href {\doibase 10.1103/physrevb.92.235119}
  {\bibfield  {journal} {\bibinfo  {journal} {Phys. Rev. B}\ }\textbf {\bibinfo
  {volume} {92}},\ \bibinfo {pages} {235119} (\bibinfo {year}
  {2015})}\BibitemShut {NoStop}%
\bibitem [{\citenamefont {Majumder}\ \emph {et~al.}(2015)\citenamefont
  {Majumder}, \citenamefont {Schmidt}, \citenamefont {Rosner}, \citenamefont
  {Tsirlin}, \citenamefont {Yasuoka},\ and\ \citenamefont
  {Baenitz}}]{Majumder2015}%
  \BibitemOpen
  \bibfield  {author} {\bibinfo {author} {\bibfnamefont {M.}~\bibnamefont
  {Majumder}}, \bibinfo {author} {\bibfnamefont {M.}~\bibnamefont {Schmidt}},
  \bibinfo {author} {\bibfnamefont {H.}~\bibnamefont {Rosner}}, \bibinfo
  {author} {\bibfnamefont {A.~A.}\ \bibnamefont {Tsirlin}}, \bibinfo {author}
  {\bibfnamefont {H.}~\bibnamefont {Yasuoka}}, \ and\ \bibinfo {author}
  {\bibfnamefont {M.}~\bibnamefont {Baenitz}},\ }\href {\doibase
  10.1103/physrevb.91.180401} {\bibfield  {journal} {\bibinfo  {journal} {Phys.
  Rev. B}\ }\textbf {\bibinfo {volume} {91}},\ \bibinfo {pages} {180401}
  (\bibinfo {year} {2015})}\BibitemShut {NoStop}%
\bibitem [{\citenamefont {Kubota}\ \emph {et~al.}(2015)\citenamefont {Kubota},
  \citenamefont {Tanaka}, \citenamefont {Ono}, \citenamefont {Narumi},\ and\
  \citenamefont {Kindo}}]{Kubota2015}%
  \BibitemOpen
  \bibfield  {author} {\bibinfo {author} {\bibfnamefont {Y.}~\bibnamefont
  {Kubota}}, \bibinfo {author} {\bibfnamefont {H.}~\bibnamefont {Tanaka}},
  \bibinfo {author} {\bibfnamefont {T.}~\bibnamefont {Ono}}, \bibinfo {author}
  {\bibfnamefont {Y.}~\bibnamefont {Narumi}}, \ and\ \bibinfo {author}
  {\bibfnamefont {K.}~\bibnamefont {Kindo}},\ }\href {\doibase
  10.1103/physrevb.91.094422} {\bibfield  {journal} {\bibinfo  {journal} {Phys.
  Rev. B}\ }\textbf {\bibinfo {volume} {91}},\ \bibinfo {pages} {094422}
  (\bibinfo {year} {2015})}\BibitemShut {NoStop}%
\bibitem [{\citenamefont {Sinn}\ \emph {et~al.}(2016)\citenamefont {Sinn},
  \citenamefont {Kim}, \citenamefont {Kim}, \citenamefont {Lee}, \citenamefont
  {Won}, \citenamefont {Oh}, \citenamefont {Han}, \citenamefont {Chang},
  \citenamefont {Hur}, \citenamefont {Sato}, \citenamefont {Park},
  \citenamefont {Kim}, \citenamefont {Kim},\ and\ \citenamefont
  {Noh}}]{Sinn2016}%
  \BibitemOpen
  \bibfield  {author} {\bibinfo {author} {\bibfnamefont {S.}~\bibnamefont
  {Sinn}}, \bibinfo {author} {\bibfnamefont {C.~H.}\ \bibnamefont {Kim}},
  \bibinfo {author} {\bibfnamefont {B.~H.}\ \bibnamefont {Kim}}, \bibinfo
  {author} {\bibfnamefont {K.~D.}\ \bibnamefont {Lee}}, \bibinfo {author}
  {\bibfnamefont {C.~J.}\ \bibnamefont {Won}}, \bibinfo {author} {\bibfnamefont
  {J.~S.}\ \bibnamefont {Oh}}, \bibinfo {author} {\bibfnamefont
  {M.}~\bibnamefont {Han}}, \bibinfo {author} {\bibfnamefont {Y.~J.}\
  \bibnamefont {Chang}}, \bibinfo {author} {\bibfnamefont {N.}~\bibnamefont
  {Hur}}, \bibinfo {author} {\bibfnamefont {H.}~\bibnamefont {Sato}}, \bibinfo
  {author} {\bibfnamefont {B.-G.}\ \bibnamefont {Park}}, \bibinfo {author}
  {\bibfnamefont {C.}~\bibnamefont {Kim}}, \bibinfo {author} {\bibfnamefont
  {H.-D.}\ \bibnamefont {Kim}}, \ and\ \bibinfo {author} {\bibfnamefont
  {T.~W.}\ \bibnamefont {Noh}},\ }\href {\doibase 10.1038/srep39544} {\bibfield
   {journal} {\bibinfo  {journal} {Sci. Rep.}\ }\textbf {\bibinfo {volume}
  {6}},\ \bibinfo {pages} {39544} (\bibinfo {year} {2016})}\BibitemShut
  {NoStop}%
\bibitem [{\citenamefont {Ziatdinov}\ \emph {et~al.}(2016)\citenamefont
  {Ziatdinov}, \citenamefont {Banerjee}, \citenamefont {Maksov}, \citenamefont
  {Berlijn}, \citenamefont {Zhou}, \citenamefont {Cao}, \citenamefont {Yan},
  \citenamefont {Bridges}, \citenamefont {Mandrus}, \citenamefont {Nagler},
  \citenamefont {Baddorf},\ and\ \citenamefont {Kalinin}}]{Ziatdinov2016}%
  \BibitemOpen
  \bibfield  {author} {\bibinfo {author} {\bibfnamefont {M.}~\bibnamefont
  {Ziatdinov}}, \bibinfo {author} {\bibfnamefont {A.}~\bibnamefont {Banerjee}},
  \bibinfo {author} {\bibfnamefont {A.}~\bibnamefont {Maksov}}, \bibinfo
  {author} {\bibfnamefont {T.}~\bibnamefont {Berlijn}}, \bibinfo {author}
  {\bibfnamefont {W.}~\bibnamefont {Zhou}}, \bibinfo {author} {\bibfnamefont
  {H.~B.}\ \bibnamefont {Cao}}, \bibinfo {author} {\bibfnamefont {J.-Q.}\
  \bibnamefont {Yan}}, \bibinfo {author} {\bibfnamefont {C.~A.}\ \bibnamefont
  {Bridges}}, \bibinfo {author} {\bibfnamefont {D.~G.}\ \bibnamefont
  {Mandrus}}, \bibinfo {author} {\bibfnamefont {S.~E.}\ \bibnamefont {Nagler}},
  \bibinfo {author} {\bibfnamefont {A.~P.}\ \bibnamefont {Baddorf}}, \ and\
  \bibinfo {author} {\bibfnamefont {S.~V.}\ \bibnamefont {Kalinin}},\ }\href
  {\doibase 10.1038/ncomms13774} {\bibfield  {journal} {\bibinfo  {journal}
  {Nat. Commun.}\ }\textbf {\bibinfo {volume} {7}},\ \bibinfo {pages} {13774}
  (\bibinfo {year} {2016})}\BibitemShut {NoStop}%
\bibitem [{\citenamefont {Weber}\ \emph {et~al.}(2016)\citenamefont {Weber},
  \citenamefont {Schoop}, \citenamefont {Duppel}, \citenamefont {Lippmann},
  \citenamefont {Nuss},\ and\ \citenamefont {Lotsch}}]{Weber2016}%
  \BibitemOpen
  \bibfield  {author} {\bibinfo {author} {\bibfnamefont {D.}~\bibnamefont
  {Weber}}, \bibinfo {author} {\bibfnamefont {L.~M.}\ \bibnamefont {Schoop}},
  \bibinfo {author} {\bibfnamefont {V.}~\bibnamefont {Duppel}}, \bibinfo
  {author} {\bibfnamefont {J.~M.}\ \bibnamefont {Lippmann}}, \bibinfo {author}
  {\bibfnamefont {J.}~\bibnamefont {Nuss}}, \ and\ \bibinfo {author}
  {\bibfnamefont {B.~V.}\ \bibnamefont {Lotsch}},\ }\href {\doibase
  10.1021/acs.nanolett.6b00701} {\bibfield  {journal} {\bibinfo  {journal}
  {Nano Lett.}\ }\textbf {\bibinfo {volume} {16}},\ \bibinfo {pages} {3578}
  (\bibinfo {year} {2016})}\BibitemShut {NoStop}%
\bibitem [{\citenamefont {Jackeli}\ and\ \citenamefont
  {Khaliullin}(2009)}]{Jackeli2009}%
  \BibitemOpen
  \bibfield  {author} {\bibinfo {author} {\bibfnamefont {G.}~\bibnamefont
  {Jackeli}}\ and\ \bibinfo {author} {\bibfnamefont {G.}~\bibnamefont
  {Khaliullin}},\ }\href {\doibase 10.1103/physrevlett.102.017205} {\bibfield
  {journal} {\bibinfo  {journal} {Phys. Rev. Lett.}\ }\textbf {\bibinfo
  {volume} {102}},\ \bibinfo {pages} {017205} (\bibinfo {year}
  {2009})}\BibitemShut {NoStop}%
\bibitem [{\citenamefont {Chaloupka}\ \emph {et~al.}(2010)\citenamefont
  {Chaloupka}, \citenamefont {Jackeli},\ and\ \citenamefont
  {Khaliullin}}]{Chaloupka2010}%
  \BibitemOpen
  \bibfield  {author} {\bibinfo {author} {\bibfnamefont {J.}~\bibnamefont
  {Chaloupka}}, \bibinfo {author} {\bibfnamefont {G.}~\bibnamefont {Jackeli}},
  \ and\ \bibinfo {author} {\bibfnamefont {G.}~\bibnamefont {Khaliullin}},\
  }\href {\doibase 10.1103/physrevlett.105.027204} {\bibfield  {journal}
  {\bibinfo  {journal} {Phys. Rev. Lett.}\ }\textbf {\bibinfo {volume} {105}},\
  \bibinfo {pages} {027204} (\bibinfo {year} {2010})}\BibitemShut {NoStop}%
\bibitem [{\citenamefont {Nussnov}\ and\ \citenamefont {van~den
  Brink}(2015)}]{Nussinov_RMP}%
  \BibitemOpen
  \bibfield  {author} {\bibinfo {author} {\bibfnamefont {Z.}~\bibnamefont
  {Nussnov}}\ and\ \bibinfo {author} {\bibfnamefont {J.}~\bibnamefont {van~den
  Brink}},\ }\href@noop {} {\bibfield  {journal} {\bibinfo  {journal} {Rev.
  Mod. Phys.}\ }\textbf {\bibinfo {volume} {87}},\ \bibinfo {pages} {1}
  (\bibinfo {year} {2015})}\BibitemShut {NoStop}%
\bibitem [{\citenamefont {Kitaev}(2006)}]{Kitaev2006}%
  \BibitemOpen
  \bibfield  {author} {\bibinfo {author} {\bibfnamefont {A.}~\bibnamefont
  {Kitaev}},\ }\href {\doibase 10.1016/j.aop.2005.10.005} {\bibfield  {journal}
  {\bibinfo  {journal} {Annals of Physics}\ }\textbf {\bibinfo {volume}
  {321}},\ \bibinfo {pages} {2} (\bibinfo {year} {2006})}\BibitemShut {NoStop}%
\bibitem [{\citenamefont {Sandilands}\ \emph {et~al.}(2015)\citenamefont
  {Sandilands}, \citenamefont {Tian}, \citenamefont {Plumb}, \citenamefont
  {Kim},\ and\ \citenamefont {Burch}}]{Sandilands2015}%
  \BibitemOpen
  \bibfield  {author} {\bibinfo {author} {\bibfnamefont {L.~J.}\ \bibnamefont
  {Sandilands}}, \bibinfo {author} {\bibfnamefont {Y.}~\bibnamefont {Tian}},
  \bibinfo {author} {\bibfnamefont {K.~W.}\ \bibnamefont {Plumb}}, \bibinfo
  {author} {\bibfnamefont {Y.-J.}\ \bibnamefont {Kim}}, \ and\ \bibinfo
  {author} {\bibfnamefont {K.~S.}\ \bibnamefont {Burch}},\ }\href {\doibase
  10.1103/physrevlett.114.147201} {\bibfield  {journal} {\bibinfo  {journal}
  {Phys. Rev. Lett.}\ }\textbf {\bibinfo {volume} {114}},\ \bibinfo {pages}
  {147201} (\bibinfo {year} {2015})}\BibitemShut {NoStop}%
\bibitem [{\citenamefont {Nasu}\ \emph {et~al.}(2015)\citenamefont {Nasu},
  \citenamefont {Udagawa},\ and\ \citenamefont {Motome}}]{Nasu2015}%
  \BibitemOpen
  \bibfield  {author} {\bibinfo {author} {\bibfnamefont {J.}~\bibnamefont
  {Nasu}}, \bibinfo {author} {\bibfnamefont {M.}~\bibnamefont {Udagawa}}, \
  and\ \bibinfo {author} {\bibfnamefont {Y.}~\bibnamefont {Motome}},\ }\href
  {\doibase 10.1103/physrevb.92.115122} {\bibfield  {journal} {\bibinfo
  {journal} {Phys. Rev. B}\ }\textbf {\bibinfo {volume} {92}},\ \bibinfo
  {pages} {115122} (\bibinfo {year} {2015})}\BibitemShut {NoStop}%
\bibitem [{\citenamefont {Trebst}()}]{Trebst2017}%
  \BibitemOpen
  \bibfield  {author} {\bibinfo {author} {\bibfnamefont {S.}~\bibnamefont
  {Trebst}},\ }\href {https://arxiv.org/abs/1701.07056} {\bibinfo  {journal}
  {arXiv:1701.07056}\ }\BibitemShut {NoStop}%
\bibitem [{\citenamefont {Singh}\ \emph {et~al.}(2012)\citenamefont {Singh},
  \citenamefont {Manni}, \citenamefont {Reuther}, \citenamefont {Berlijn},
  \citenamefont {Thomale}, \citenamefont {Ku}, \citenamefont {Trebst},\ and\
  \citenamefont {Gegenwart}}]{Singh2012}%
  \BibitemOpen
\bibfield  {journal} {  }\bibfield  {author} {\bibinfo {author} {\bibfnamefont
  {Y.}~\bibnamefont {Singh}}, \bibinfo {author} {\bibfnamefont
  {S.}~\bibnamefont {Manni}}, \bibinfo {author} {\bibfnamefont
  {J.}~\bibnamefont {Reuther}}, \bibinfo {author} {\bibfnamefont
  {T.}~\bibnamefont {Berlijn}}, \bibinfo {author} {\bibfnamefont
  {R.}~\bibnamefont {Thomale}}, \bibinfo {author} {\bibfnamefont
  {W.}~\bibnamefont {Ku}}, \bibinfo {author} {\bibfnamefont {S.}~\bibnamefont
  {Trebst}}, \ and\ \bibinfo {author} {\bibfnamefont {P.}~\bibnamefont
  {Gegenwart}},\ }\href {\doibase 10.1103/physrevlett.108.127203} {\bibfield
  {journal} {\bibinfo  {journal} {Phys. Rev. Lett.}\ }\textbf {\bibinfo
  {volume} {108}},\ \bibinfo {pages} {127203} (\bibinfo {year}
  {2012})}\BibitemShut {NoStop}%
\bibitem [{\citenamefont {Banerjee}\ \emph {et~al.}(2016)\citenamefont
  {Banerjee}, \citenamefont {Bridges}, \citenamefont {Yan}, \citenamefont
  {Aczel}, \citenamefont {Li}, \citenamefont {Stone}, \citenamefont {Granroth},
  \citenamefont {Lumsden}, \citenamefont {Yiu}, \citenamefont {Knolle},
  \citenamefont {Bhattacharjee}, \citenamefont {Kovrizhin}, \citenamefont
  {Moessner}, \citenamefont {Tennant}, \citenamefont {Mandrus},\ and\
  \citenamefont {Nagler}}]{Banerjee2016}%
  \BibitemOpen
  \bibfield  {author} {\bibinfo {author} {\bibfnamefont {A.}~\bibnamefont
  {Banerjee}}, \bibinfo {author} {\bibfnamefont {C.~A.}\ \bibnamefont
  {Bridges}}, \bibinfo {author} {\bibfnamefont {J.-Q.}\ \bibnamefont {Yan}},
  \bibinfo {author} {\bibfnamefont {A.~A.}\ \bibnamefont {Aczel}}, \bibinfo
  {author} {\bibfnamefont {L.}~\bibnamefont {Li}}, \bibinfo {author}
  {\bibfnamefont {M.~B.}\ \bibnamefont {Stone}}, \bibinfo {author}
  {\bibfnamefont {G.~E.}\ \bibnamefont {Granroth}}, \bibinfo {author}
  {\bibfnamefont {M.~D.}\ \bibnamefont {Lumsden}}, \bibinfo {author}
  {\bibfnamefont {Y.}~\bibnamefont {Yiu}}, \bibinfo {author} {\bibfnamefont
  {J.}~\bibnamefont {Knolle}}, \bibinfo {author} {\bibfnamefont
  {S.}~\bibnamefont {Bhattacharjee}}, \bibinfo {author} {\bibfnamefont {D.~L.}\
  \bibnamefont {Kovrizhin}}, \bibinfo {author} {\bibfnamefont {R.}~\bibnamefont
  {Moessner}}, \bibinfo {author} {\bibfnamefont {D.~A.}\ \bibnamefont
  {Tennant}}, \bibinfo {author} {\bibfnamefont {D.~G.}\ \bibnamefont
  {Mandrus}}, \ and\ \bibinfo {author} {\bibfnamefont {S.~E.}\ \bibnamefont
  {Nagler}},\ }\href {\doibase 10.1038/nmat4604} {\bibfield  {journal}
  {\bibinfo  {journal} {Nat. Mater.}\ }\textbf {\bibinfo {volume} {15}},\
  \bibinfo {pages} {733} (\bibinfo {year} {2016})}\BibitemShut {NoStop}%
\bibitem [{\citenamefont {Cao}\ \emph {et~al.}(2016)\citenamefont {Cao},
  \citenamefont {Banerjee}, \citenamefont {Yan}, \citenamefont {Bridges},
  \citenamefont {Lumsden}, \citenamefont {Mandrus}, \citenamefont {Tennant},
  \citenamefont {Chakoumakos},\ and\ \citenamefont {Nagler}}]{Cao2016}%
  \BibitemOpen
  \bibfield  {author} {\bibinfo {author} {\bibfnamefont {H.~B.}\ \bibnamefont
  {Cao}}, \bibinfo {author} {\bibfnamefont {A.}~\bibnamefont {Banerjee}},
  \bibinfo {author} {\bibfnamefont {J.-Q.}\ \bibnamefont {Yan}}, \bibinfo
  {author} {\bibfnamefont {C.~A.}\ \bibnamefont {Bridges}}, \bibinfo {author}
  {\bibfnamefont {M.~D.}\ \bibnamefont {Lumsden}}, \bibinfo {author}
  {\bibfnamefont {D.~G.}\ \bibnamefont {Mandrus}}, \bibinfo {author}
  {\bibfnamefont {D.~A.}\ \bibnamefont {Tennant}}, \bibinfo {author}
  {\bibfnamefont {B.~C.}\ \bibnamefont {Chakoumakos}}, \ and\ \bibinfo {author}
  {\bibfnamefont {S.~E.}\ \bibnamefont {Nagler}},\ }\href {\doibase
  10.1103/physrevb.93.134423} {\bibfield  {journal} {\bibinfo  {journal} {Phys.
  Rev. B}\ }\textbf {\bibinfo {volume} {93}},\ \bibinfo {pages} {134423}
  (\bibinfo {year} {2016})}\BibitemShut {NoStop}%
\bibitem [{\citenamefont {Yadav}\ \emph {et~al.}(2016)\citenamefont {Yadav},
  \citenamefont {Bogdanov}, \citenamefont {Katukuri}, \citenamefont
  {Nishimoto}, \citenamefont {van~den Brink},\ and\ \citenamefont
  {Hozoi}}]{Yadav2016}%
  \BibitemOpen
  \bibfield  {author} {\bibinfo {author} {\bibfnamefont {R.}~\bibnamefont
  {Yadav}}, \bibinfo {author} {\bibfnamefont {N.~A.}\ \bibnamefont {Bogdanov}},
  \bibinfo {author} {\bibfnamefont {V.~M.}\ \bibnamefont {Katukuri}}, \bibinfo
  {author} {\bibfnamefont {S.}~\bibnamefont {Nishimoto}}, \bibinfo {author}
  {\bibfnamefont {J.}~\bibnamefont {van~den Brink}}, \ and\ \bibinfo {author}
  {\bibfnamefont {L.}~\bibnamefont {Hozoi}},\ }\href {\doibase
  10.1038/srep37925} {\bibfield  {journal} {\bibinfo  {journal} {Sci. Rep.}\
  }\textbf {\bibinfo {volume} {6}},\ \bibinfo {pages} {37925} (\bibinfo {year}
  {2016})}\BibitemShut {NoStop}%
\bibitem [{\citenamefont {Winter}\ \emph {et~al.}(2016)\citenamefont {Winter},
  \citenamefont {Li}, \citenamefont {Jeschke},\ and\ \citenamefont
  {Valent{\'{\i}}}}]{Winter2016}%
  \BibitemOpen
  \bibfield  {author} {\bibinfo {author} {\bibfnamefont {S.~M.}\ \bibnamefont
  {Winter}}, \bibinfo {author} {\bibfnamefont {Y.}~\bibnamefont {Li}}, \bibinfo
  {author} {\bibfnamefont {H.~O.}\ \bibnamefont {Jeschke}}, \ and\ \bibinfo
  {author} {\bibfnamefont {R.}~\bibnamefont {Valent{\'{\i}}}},\ }\href
  {\doibase 10.1103/physrevb.93.214431} {\bibfield  {journal} {\bibinfo
  {journal} {Phys. Rev. B}\ }\textbf {\bibinfo {volume} {93}},\ \bibinfo
  {pages} {214431} (\bibinfo {year} {2016})}\BibitemShut {NoStop}%
\bibitem [{\citenamefont {Kimchi}\ and\ \citenamefont
  {You}(2011)}]{Kimchi2011}%
  \BibitemOpen
  \bibfield  {author} {\bibinfo {author} {\bibfnamefont {I.}~\bibnamefont
  {Kimchi}}\ and\ \bibinfo {author} {\bibfnamefont {Y.-Z.}\ \bibnamefont
  {You}},\ }\href {\doibase 10.1103/physrevb.84.180407} {\bibfield  {journal}
  {\bibinfo  {journal} {Phys. Rev. B}\ }\textbf {\bibinfo {volume} {84}},\
  \bibinfo {pages} {180407} (\bibinfo {year} {2011})}\BibitemShut {NoStop}%
\bibitem [{\citenamefont {Rau}\ \emph {et~al.}(2014)\citenamefont {Rau},
  \citenamefont {Lee},\ and\ \citenamefont {Kee}}]{Rau2014}%
  \BibitemOpen
  \bibfield  {author} {\bibinfo {author} {\bibfnamefont {J.~G.}\ \bibnamefont
  {Rau}}, \bibinfo {author} {\bibfnamefont {E.~K.-H.}\ \bibnamefont {Lee}}, \
  and\ \bibinfo {author} {\bibfnamefont {H.-Y.}\ \bibnamefont {Kee}},\ }\href
  {\doibase 10.1103/PhysRevLett.112.077204} {\bibfield  {journal} {\bibinfo
  {journal} {Phys. Rev. Lett.}\ }\textbf {\bibinfo {volume} {112}},\ \bibinfo
  {pages} {077204} (\bibinfo {year} {2014})}\BibitemShut {NoStop}%
\bibitem [{\citenamefont {Yamaji}\ \emph {et~al.}(2014)\citenamefont {Yamaji},
  \citenamefont {Nomura}, \citenamefont {Kurita}, \citenamefont {Arita},\ and\
  \citenamefont {Imada}}]{Yamaji2014}%
  \BibitemOpen
  \bibfield  {author} {\bibinfo {author} {\bibfnamefont {Y.}~\bibnamefont
  {Yamaji}}, \bibinfo {author} {\bibfnamefont {Y.}~\bibnamefont {Nomura}},
  \bibinfo {author} {\bibfnamefont {M.}~\bibnamefont {Kurita}}, \bibinfo
  {author} {\bibfnamefont {R.}~\bibnamefont {Arita}}, \ and\ \bibinfo {author}
  {\bibfnamefont {M.}~\bibnamefont {Imada}},\ }\href {\doibase
  10.1103/physrevlett.113.107201} {\bibfield  {journal} {\bibinfo  {journal}
  {Phys. Rev. Lett.}\ }\textbf {\bibinfo {volume} {113}},\ \bibinfo {pages}
  {107201} (\bibinfo {year} {2014})}\BibitemShut {NoStop}%
\bibitem [{\citenamefont {Reuther}\ \emph {et~al.}(2014)\citenamefont
  {Reuther}, \citenamefont {Thomale},\ and\ \citenamefont
  {Rachel}}]{Reuther2014}%
  \BibitemOpen
  \bibfield  {author} {\bibinfo {author} {\bibfnamefont {J.}~\bibnamefont
  {Reuther}}, \bibinfo {author} {\bibfnamefont {R.}~\bibnamefont {Thomale}}, \
  and\ \bibinfo {author} {\bibfnamefont {S.}~\bibnamefont {Rachel}},\ }\href
  {\doibase 10.1103/physrevb.90.100405} {\bibfield  {journal} {\bibinfo
  {journal} {Phys. Rev. B}\ }\textbf {\bibinfo {volume} {90}},\ \bibinfo
  {pages} {100405} (\bibinfo {year} {2014})}\BibitemShut {NoStop}%
\bibitem [{\citenamefont {Sizyuk}\ \emph {et~al.}(2014)\citenamefont {Sizyuk},
  \citenamefont {Price}, \citenamefont {W\"olfle},\ and\ \citenamefont
  {Perkins}}]{Perkins2014}%
  \BibitemOpen
  \bibfield  {author} {\bibinfo {author} {\bibfnamefont {Y.}~\bibnamefont
  {Sizyuk}}, \bibinfo {author} {\bibfnamefont {C.}~\bibnamefont {Price}},
  \bibinfo {author} {\bibfnamefont {P.}~\bibnamefont {W\"olfle}}, \ and\
  \bibinfo {author} {\bibfnamefont {N.~B.}\ \bibnamefont {Perkins}},\
  }\href@noop {} {\bibfield  {journal} {\bibinfo  {journal} {Phys. Rev. B}\
  }\textbf {\bibinfo {volume} {90}},\ \bibinfo {pages} {155126} (\bibinfo
  {year} {2014})}\BibitemShut {NoStop}%
\bibitem [{\citenamefont {Rousochatzakis}\ \emph {et~al.}(2015)\citenamefont
  {Rousochatzakis}, \citenamefont {Reuther}, \citenamefont {Thomale},
  \citenamefont {Rachel},\ and\ \citenamefont {Perkins}}]{Rousso2015}%
  \BibitemOpen
  \bibfield  {author} {\bibinfo {author} {\bibfnamefont {I.}~\bibnamefont
  {Rousochatzakis}}, \bibinfo {author} {\bibfnamefont {J.}~\bibnamefont
  {Reuther}}, \bibinfo {author} {\bibfnamefont {R.}~\bibnamefont {Thomale}},
  \bibinfo {author} {\bibfnamefont {S.}~\bibnamefont {Rachel}}, \ and\ \bibinfo
  {author} {\bibfnamefont {N.~B.}\ \bibnamefont {Perkins}},\ }\href@noop {}
  {\bibfield  {journal} {\bibinfo  {journal} {Phys. Rev. X}\ }\textbf {\bibinfo
  {volume} {5}},\ \bibinfo {pages} {041035} (\bibinfo {year}
  {2015})}\BibitemShut {NoStop}%
\bibitem [{\citenamefont {Winter}\ \emph {et~al.}()\citenamefont {Winter},
  \citenamefont {Riedl}, \citenamefont {Honecker},\ and\ \citenamefont
  {Valent{\'{\i}}}}]{Winter2017}%
  \BibitemOpen
  \bibfield  {author} {\bibinfo {author} {\bibfnamefont {S.~M.}\ \bibnamefont
  {Winter}}, \bibinfo {author} {\bibfnamefont {K.}~\bibnamefont {Riedl}},
  \bibinfo {author} {\bibfnamefont {A.}~\bibnamefont {Honecker}}, \ and\
  \bibinfo {author} {\bibfnamefont {R.}~\bibnamefont {Valent{\'{\i}}}},\
  }\href@noop {} {\bibinfo  {journal} {arXiv:1702.08466}\ }\BibitemShut
  {NoStop}%
\bibitem [{\citenamefont {Laubach}\ \emph {et~al.}()\citenamefont {Laubach},
  \citenamefont {Reuther}, \citenamefont {Thomale},\ and\ \citenamefont
  {Rachel}}]{Laubach2017}%
  \BibitemOpen
\bibfield  {journal} {  }\bibfield  {author} {\bibinfo {author} {\bibfnamefont
  {M.}~\bibnamefont {Laubach}}, \bibinfo {author} {\bibfnamefont
  {J.}~\bibnamefont {Reuther}}, \bibinfo {author} {\bibfnamefont
  {R.}~\bibnamefont {Thomale}}, \ and\ \bibinfo {author} {\bibfnamefont
  {S.}~\bibnamefont {Rachel}},\ }\href@noop {} {\bibinfo  {journal}
  {arXiv:1701.04896}\ }\BibitemShut {NoStop}%
\bibitem [{\citenamefont {Leahy}\ \emph {et~al.}()\citenamefont {Leahy},
  \citenamefont {Pocs}, \citenamefont {Siegfried}, \citenamefont {anbd
  S.-H.~Do}, \citenamefont {Choi}, \citenamefont {Normand},\ and\ \citenamefont
  {Lee}}]{Leahy2016}%
  \BibitemOpen
\bibfield  {journal} {  }\bibfield  {author} {\bibinfo {author} {\bibfnamefont
  {I.~A.}\ \bibnamefont {Leahy}}, \bibinfo {author} {\bibfnamefont {C.~A.}\
  \bibnamefont {Pocs}}, \bibinfo {author} {\bibfnamefont {P.~E.}\ \bibnamefont
  {Siegfried}}, \bibinfo {author} {\bibfnamefont {D.~G.}\ \bibnamefont {anbd
  S.-H.~Do}}, \bibinfo {author} {\bibfnamefont {K.-Y.}\ \bibnamefont {Choi}},
  \bibinfo {author} {\bibfnamefont {B.}~\bibnamefont {Normand}}, \ and\
  \bibinfo {author} {\bibfnamefont {M.}~\bibnamefont {Lee}},\ }\href@noop {}
  {\bibinfo  {journal} {arXiv:1612.03881}\ }\BibitemShut {NoStop}%
\bibitem [{\citenamefont {Baek}\ \emph {et~al.}()\citenamefont {Baek},
  \citenamefont {Do}, \citenamefont {Choi}, \citenamefont {Kwon}, \citenamefont
  {Wolter}, \citenamefont {Nishimoto}, \citenamefont {van~den Brink},\ and\
  \citenamefont {B\"uchner}}]{Baek2017}%
  \BibitemOpen
\bibfield  {journal} {  }\bibfield  {author} {\bibinfo {author} {\bibfnamefont
  {S.-H.}\ \bibnamefont {Baek}}, \bibinfo {author} {\bibfnamefont {S.-H.}\
  \bibnamefont {Do}}, \bibinfo {author} {\bibfnamefont {K.-Y.}\ \bibnamefont
  {Choi}}, \bibinfo {author} {\bibfnamefont {Y.~S.}\ \bibnamefont {Kwon}},
  \bibinfo {author} {\bibfnamefont {A.~U.~B.}\ \bibnamefont {Wolter}}, \bibinfo
  {author} {\bibfnamefont {S.}~\bibnamefont {Nishimoto}}, \bibinfo {author}
  {\bibfnamefont {J.}~\bibnamefont {van~den Brink}}, \ and\ \bibinfo {author}
  {\bibfnamefont {B.}~\bibnamefont {B\"uchner}},\ }\href@noop {} {\bibinfo
  {journal} {arXiv:1702.01671}\ }\BibitemShut {NoStop}%
\bibitem [{sup()}]{suppl}%
  \BibitemOpen
\bibfield  {journal} {  }\href@noop {} {}\bibinfo {note} {See Supplemental
  Material for a magnetic characterization of the sample, the phonon
  calculations for RhCl$_3$, and for a semiclassical analysis of a {\JKG} model
  in a magnetic field.}\BibitemShut {Stop}%
\bibitem [{\citenamefont {Lindemann}(1910)}]{Lindemann1910}%
  \BibitemOpen
  \bibfield  {author} {\bibinfo {author} {\bibfnamefont {F.~A.}\ \bibnamefont
  {Lindemann}},\ }\href@noop {} {\bibfield  {journal} {\bibinfo  {journal}
  {Physik. Z.}\ }\textbf {\bibinfo {volume} {11}},\ \bibinfo {pages} {609}
  (\bibinfo {year} {1910})}\BibitemShut {NoStop}%
\bibitem [{\citenamefont {Sachdev}(2011)}]{ssbook}%
  \BibitemOpen
  \bibfield  {author} {\bibinfo {author} {\bibfnamefont {S.}~\bibnamefont
  {Sachdev}},\ }\href@noop {} {\emph {\bibinfo {title} {{Quantum Phase
  Transitions}}}},\ \bibinfo {edition} {2nd}\ ed.\ (\bibinfo  {publisher}
  {Cambridge University Press},\ \bibinfo {year} {2011})\BibitemShut {NoStop}%
\bibitem [{\citenamefont {Vojta}(2003)}]{Vojta2003}%
  \BibitemOpen
  \bibfield  {author} {\bibinfo {author} {\bibfnamefont {M.}~\bibnamefont
  {Vojta}},\ }\href {\doibase 10.1088/0034-4885/66/12/r01} {\bibfield
  {journal} {\bibinfo  {journal} {Rep. Prog. Phys.}\ }\textbf {\bibinfo
  {volume} {66}},\ \bibinfo {pages} {2069} (\bibinfo {year}
  {2003})}\BibitemShut {NoStop}%
\bibitem [{\citenamefont {Janssen}\ \emph {et~al.}(2016)\citenamefont
  {Janssen}, \citenamefont {Andrade},\ and\ \citenamefont
  {Vojta}}]{Janssen2016}%
  \BibitemOpen
  \bibfield  {author} {\bibinfo {author} {\bibfnamefont {L.}~\bibnamefont
  {Janssen}}, \bibinfo {author} {\bibfnamefont {E.~C.}\ \bibnamefont
  {Andrade}}, \ and\ \bibinfo {author} {\bibfnamefont {M.}~\bibnamefont
  {Vojta}},\ }\href@noop {} {\bibfield  {journal} {\bibinfo  {journal} {Phys.
  Rev. Lett.}\ }\textbf {\bibinfo {volume} {117}},\ \bibinfo {pages} {277202}
  (\bibinfo {year} {2016})}\BibitemShut {NoStop}%
\bibitem [{\citenamefont {Gohlke}\ \emph {et~al.}()\citenamefont {Gohlke},
  \citenamefont {Verresen}, \citenamefont {Moessner},\ and\ \citenamefont
  {Pollmann}}]{Gohlke2017}%
  \BibitemOpen
  \bibfield  {author} {\bibinfo {author} {\bibfnamefont {M.}~\bibnamefont
  {Gohlke}}, \bibinfo {author} {\bibfnamefont {R.}~\bibnamefont {Verresen}},
  \bibinfo {author} {\bibfnamefont {R.}~\bibnamefont {Moessner}}, \ and\
  \bibinfo {author} {\bibfnamefont {F.}~\bibnamefont {Pollmann}},\ }\href@noop
  {} {\bibinfo  {journal} {arXiv:1701.04678}\ }\BibitemShut {NoStop}%
\bibitem [{\citenamefont {Sears}\ \emph {et~al.}()\citenamefont {Sears},
  \citenamefont {Zhao}, \citenamefont {Xu}, \citenamefont {Lynn},\ and\
  \citenamefont {Kim}}]{Sears2017}%
  \BibitemOpen
\bibfield  {journal} {  }\bibfield  {author} {\bibinfo {author} {\bibfnamefont
  {J.~A.}\ \bibnamefont {Sears}}, \bibinfo {author} {\bibfnamefont
  {Y.}~\bibnamefont {Zhao}}, \bibinfo {author} {\bibfnamefont {Z.}~\bibnamefont
  {Xu}}, \bibinfo {author} {\bibfnamefont {J.~W.}\ \bibnamefont {Lynn}}, \ and\
  \bibinfo {author} {\bibfnamefont {Y.-J.}\ \bibnamefont {Kim}},\ }\href@noop
  {} {\bibinfo  {journal} {arXiv:1703.08431}\ }\BibitemShut {NoStop}%
\bibitem [{\citenamefont {Zheng}\ \emph {et~al.}()\citenamefont {Zheng},
  \citenamefont {Ran}, \citenamefont {Li}, \citenamefont {Wang}, \citenamefont
  {Wang}, \citenamefont {Liu}, \citenamefont {Liu}, \citenamefont {Normand},
  \citenamefont {Wen},\ and\ \citenamefont {Yu}}]{Zheng2017}%
  \BibitemOpen
\bibfield  {journal} {  }\bibfield  {author} {\bibinfo {author} {\bibfnamefont
  {J.}~\bibnamefont {Zheng}}, \bibinfo {author} {\bibfnamefont
  {K.}~\bibnamefont {Ran}}, \bibinfo {author} {\bibfnamefont {T.}~\bibnamefont
  {Li}}, \bibinfo {author} {\bibfnamefont {J.}~\bibnamefont {Wang}}, \bibinfo
  {author} {\bibfnamefont {P.}~\bibnamefont {Wang}}, \bibinfo {author}
  {\bibfnamefont {B.}~\bibnamefont {Liu}}, \bibinfo {author} {\bibfnamefont
  {Z.}~\bibnamefont {Liu}}, \bibinfo {author} {\bibfnamefont {B.}~\bibnamefont
  {Normand}}, \bibinfo {author} {\bibfnamefont {J.}~\bibnamefont {Wen}}, \ and\
  \bibinfo {author} {\bibfnamefont {W.}~\bibnamefont {Yu}},\ }\href@noop {}
  {\bibinfo  {journal} {arXiv:1703.08474}\ }\BibitemShut {NoStop}%
\end{thebibliography}%


\begin{thebibliography}{18}%
\makeatletter
\providecommand \@ifxundefined [1]{%
 \@ifx{#1\undefined}
}%
\providecommand \@ifnum [1]{%
 \ifnum #1\expandafter \@firstoftwo
 \else \expandafter \@secondoftwo
 \fi
}%
\providecommand \@ifx [1]{%
 \ifx #1\expandafter \@firstoftwo
 \else \expandafter \@secondoftwo
 \fi
}%
\providecommand \natexlab [1]{#1}%
\providecommand \enquote  [1]{``#1''}%
\providecommand \bibnamefont  [1]{#1}%
\providecommand \bibfnamefont [1]{#1}%
\providecommand \citenamefont [1]{#1}%
\providecommand \href@noop [0]{\@secondoftwo}%
\providecommand \href [0]{\begingroup \@sanitize@url \@href}%
\providecommand \@href[1]{\@@startlink{#1}\@@href}%
\providecommand \@@href[1]{\endgroup#1\@@endlink}%
\providecommand \@sanitize@url [0]{\catcode `\\12\catcode `\$12\catcode
  `\&12\catcode `\#12\catcode `\^12\catcode `\_12\catcode `\%12\relax}%
\providecommand \@@startlink[1]{}%
\providecommand \@@endlink[0]{}%
\providecommand \url  [0]{\begingroup\@sanitize@url \@url }%
\providecommand \@url [1]{\endgroup\@href {#1}{\urlprefix }}%
\providecommand \urlprefix  [0]{URL }%
\providecommand \Eprint [0]{\href }%
\providecommand \doibase [0]{http://dx.doi.org/}%
\providecommand \selectlanguage [0]{\@gobble}%
\providecommand \bibinfo  [0]{\@secondoftwo}%
\providecommand \bibfield  [0]{\@secondoftwo}%
\providecommand \translation [1]{[#1]}%
\providecommand \BibitemOpen [0]{}%
\providecommand \bibitemStop [0]{}%
\providecommand \bibitemNoStop [0]{.\EOS\space}%
\providecommand \EOS [0]{\spacefactor3000\relax}%
\providecommand \BibitemShut  [1]{\csname bibitem#1\endcsname}%
\let\auto@bib@innerbib\@empty
%</preamble>
\bibitem [{\citenamefont {Banerjee}\ \emph {et~al.}(2016)\citenamefont
  {Banerjee}, \citenamefont {Bridges}, \citenamefont {Yan}, \citenamefont
  {Aczel}, \citenamefont {Li}, \citenamefont {Stone}, \citenamefont {Granroth},
  \citenamefont {Lumsden}, \citenamefont {Yiu}, \citenamefont {Knolle},
  \citenamefont {Bhattacharjee}, \citenamefont {Kovrizhin}, \citenamefont
  {Moessner}, \citenamefont {Tennant}, \citenamefont {Mandrus},\ and\
  \citenamefont {Nagler}}]{Banerjee2016}%
  \BibitemOpen
  \bibfield  {author} {\bibinfo {author} {\bibfnamefont {A.}~\bibnamefont
  {Banerjee}}, \bibinfo {author} {\bibfnamefont {C.~A.}\ \bibnamefont
  {Bridges}}, \bibinfo {author} {\bibfnamefont {J.-Q.}\ \bibnamefont {Yan}},
  \bibinfo {author} {\bibfnamefont {A.~A.}\ \bibnamefont {Aczel}}, \bibinfo
  {author} {\bibfnamefont {L.}~\bibnamefont {Li}}, \bibinfo {author}
  {\bibfnamefont {M.~B.}\ \bibnamefont {Stone}}, \bibinfo {author}
  {\bibfnamefont {G.~E.}\ \bibnamefont {Granroth}}, \bibinfo {author}
  {\bibfnamefont {M.~D.}\ \bibnamefont {Lumsden}}, \bibinfo {author}
  {\bibfnamefont {Y.}~\bibnamefont {Yiu}}, \bibinfo {author} {\bibfnamefont
  {J.}~\bibnamefont {Knolle}}, \bibinfo {author} {\bibfnamefont
  {S.}~\bibnamefont {Bhattacharjee}}, \bibinfo {author} {\bibfnamefont {D.~L.}\
  \bibnamefont {Kovrizhin}}, \bibinfo {author} {\bibfnamefont {R.}~\bibnamefont
  {Moessner}}, \bibinfo {author} {\bibfnamefont {D.~A.}\ \bibnamefont
  {Tennant}}, \bibinfo {author} {\bibfnamefont {D.~G.}\ \bibnamefont
  {Mandrus}}, \ and\ \bibinfo {author} {\bibfnamefont {S.~E.}\ \bibnamefont
  {Nagler}},\ }\href {\doibase 10.1038/nmat4604} {\bibfield  {journal}
  {\bibinfo  {journal} {Nat. Mater.}\ }\textbf {\bibinfo {volume} {15}},\
  \bibinfo {pages} {733} (\bibinfo {year} {2016})}\BibitemShut {NoStop}%
\bibitem [{\citenamefont {Cao}\ \emph {et~al.}(2016)\citenamefont {Cao},
  \citenamefont {Banerjee}, \citenamefont {Yan}, \citenamefont {Bridges},
  \citenamefont {Lumsden}, \citenamefont {Mandrus}, \citenamefont {Tennant},
  \citenamefont {Chakoumakos},\ and\ \citenamefont {Nagler}}]{Cao2016}%
  \BibitemOpen
  \bibfield  {author} {\bibinfo {author} {\bibfnamefont {H.~B.}\ \bibnamefont
  {Cao}}, \bibinfo {author} {\bibfnamefont {A.}~\bibnamefont {Banerjee}},
  \bibinfo {author} {\bibfnamefont {J.-Q.}\ \bibnamefont {Yan}}, \bibinfo
  {author} {\bibfnamefont {C.~A.}\ \bibnamefont {Bridges}}, \bibinfo {author}
  {\bibfnamefont {M.~D.}\ \bibnamefont {Lumsden}}, \bibinfo {author}
  {\bibfnamefont {D.~G.}\ \bibnamefont {Mandrus}}, \bibinfo {author}
  {\bibfnamefont {D.~A.}\ \bibnamefont {Tennant}}, \bibinfo {author}
  {\bibfnamefont {B.~C.}\ \bibnamefont {Chakoumakos}}, \ and\ \bibinfo {author}
  {\bibfnamefont {S.~E.}\ \bibnamefont {Nagler}},\ }\href {\doibase
  10.1103/physrevb.93.134423} {\bibfield  {journal} {\bibinfo  {journal} {Phys.
  Rev. B}\ }\textbf {\bibinfo {volume} {93}},\ \bibinfo {pages} {134423}
  (\bibinfo {year} {2016})}\BibitemShut {NoStop}%
\bibitem [{\citenamefont {Baek}\ \emph {et~al.}()\citenamefont {Baek},
  \citenamefont {Do}, \citenamefont {Choi}, \citenamefont {Kwon}, \citenamefont
  {Wolter}, \citenamefont {Nishimoto}, \citenamefont {van~den Brink},\ and\
  \citenamefont {B\"uchner}}]{Baek2017}%
  \BibitemOpen
  \bibfield  {author} {\bibinfo {author} {\bibfnamefont {S.-H.}\ \bibnamefont
  {Baek}}, \bibinfo {author} {\bibfnamefont {S.-H.}\ \bibnamefont {Do}},
  \bibinfo {author} {\bibfnamefont {K.-Y.}\ \bibnamefont {Choi}}, \bibinfo
  {author} {\bibfnamefont {Y.~S.}\ \bibnamefont {Kwon}}, \bibinfo {author}
  {\bibfnamefont {A.~U.~B.}\ \bibnamefont {Wolter}}, \bibinfo {author}
  {\bibfnamefont {S.}~\bibnamefont {Nishimoto}}, \bibinfo {author}
  {\bibfnamefont {J.}~\bibnamefont {van~den Brink}}, \ and\ \bibinfo {author}
  {\bibfnamefont {B.}~\bibnamefont {B\"uchner}},\ }\href@noop {} {\bibinfo
  {journal} {arXiv:1702.01671}\ }\BibitemShut {NoStop}%
\bibitem [{\citenamefont {Sears}\ \emph {et~al.}()\citenamefont {Sears},
  \citenamefont {Zhao}, \citenamefont {Xu}, \citenamefont {Lynn},\ and\
  \citenamefont {Kim}}]{Sears2017}%
  \BibitemOpen
\bibfield  {journal} {  }\bibfield  {author} {\bibinfo {author} {\bibfnamefont
  {J.~A.}\ \bibnamefont {Sears}}, \bibinfo {author} {\bibfnamefont
  {Y.}~\bibnamefont {Zhao}}, \bibinfo {author} {\bibfnamefont {Z.}~\bibnamefont
  {Xu}}, \bibinfo {author} {\bibfnamefont {J.~W.}\ \bibnamefont {Lynn}}, \ and\
  \bibinfo {author} {\bibfnamefont {Y.-J.}\ \bibnamefont {Kim}},\ }\href@noop
  {} {\bibinfo  {journal} {arXiv:1703.08431}\ }\BibitemShut {NoStop}%
\bibitem [{\citenamefont {Bl\"ochl}(1994)}]{Bloechl:1994}%
  \BibitemOpen
\bibfield  {journal} {  }\bibfield  {author} {\bibinfo {author} {\bibfnamefont
  {P.~E.}\ \bibnamefont {Bl\"ochl}},\ }\href {\doibase
  10.1103/PhysRevB.50.17953} {\bibfield  {journal} {\bibinfo  {journal} {Phys.
  Rev. B}\ }\textbf {\bibinfo {volume} {50}},\ \bibinfo {pages} {17953}
  (\bibinfo {year} {1994})}\BibitemShut {NoStop}%
\bibitem [{\citenamefont {Kresse}\ and\ \citenamefont
  {Joubert}(1999)}]{Kresse:1999}%
  \BibitemOpen
  \bibfield  {author} {\bibinfo {author} {\bibfnamefont {G.}~\bibnamefont
  {Kresse}}\ and\ \bibinfo {author} {\bibfnamefont {D.}~\bibnamefont
  {Joubert}},\ }\href {\doibase 10.1103/PhysRevB.59.1758} {\bibfield  {journal}
  {\bibinfo  {journal} {Phys. Rev. B}\ }\textbf {\bibinfo {volume} {59}},\
  \bibinfo {pages} {1758} (\bibinfo {year} {1999})}\BibitemShut {NoStop}%
\bibitem [{\citenamefont {Kresse}\ and\ \citenamefont
  {Furthm\"uller}(1996)}]{Kresse:1996}%
  \BibitemOpen
  \bibfield  {author} {\bibinfo {author} {\bibfnamefont {G.}~\bibnamefont
  {Kresse}}\ and\ \bibinfo {author} {\bibfnamefont {J.}~\bibnamefont
  {Furthm\"uller}},\ }\href@noop {} {\bibfield  {journal} {\bibinfo  {journal}
  {Phys. Rev. B}\ }\textbf {\bibinfo {volume} {54}},\ \bibinfo {pages} {11169}
  (\bibinfo {year} {1996})}\BibitemShut {NoStop}%
\bibitem [{\citenamefont {Parlinski}\ \emph {et~al.}(1997)\citenamefont
  {Parlinski}, \citenamefont {Li},\ and\ \citenamefont
  {Kawazoe}}]{Parlinski:1997}%
  \BibitemOpen
  \bibfield  {author} {\bibinfo {author} {\bibfnamefont {K.}~\bibnamefont
  {Parlinski}}, \bibinfo {author} {\bibfnamefont {Z.~Q.}\ \bibnamefont {Li}}, \
  and\ \bibinfo {author} {\bibfnamefont {Y.}~\bibnamefont {Kawazoe}},\ }\href
  {\doibase 10.1103/PhysRevLett.78.4063} {\bibfield  {journal} {\bibinfo
  {journal} {Phys. Rev. Lett.}\ }\textbf {\bibinfo {volume} {78}},\ \bibinfo
  {pages} {4063} (\bibinfo {year} {1997})}\BibitemShut {NoStop}%
\bibitem [{\citenamefont {Chaput}\ \emph {et~al.}(2011)\citenamefont {Chaput},
  \citenamefont {Togo}, \citenamefont {Tanaka},\ and\ \citenamefont
  {Hug}}]{Chaput:2011}%
  \BibitemOpen
  \bibfield  {author} {\bibinfo {author} {\bibfnamefont {L.}~\bibnamefont
  {Chaput}}, \bibinfo {author} {\bibfnamefont {A.}~\bibnamefont {Togo}},
  \bibinfo {author} {\bibfnamefont {I.}~\bibnamefont {Tanaka}}, \ and\ \bibinfo
  {author} {\bibfnamefont {G.}~\bibnamefont {Hug}},\ }\href {\doibase
  10.1103/PhysRevB.84.094302} {\bibfield  {journal} {\bibinfo  {journal} {Phys.
  Rev. B}\ }\textbf {\bibinfo {volume} {84}},\ \bibinfo {pages} {094302}
  (\bibinfo {year} {2011})}\BibitemShut {NoStop}%
\bibitem [{\citenamefont {Gonze}\ and\ \citenamefont
  {Lee}(1997)}]{Gonze:1997b}%
  \BibitemOpen
  \bibfield  {author} {\bibinfo {author} {\bibfnamefont {X.}~\bibnamefont
  {Gonze}}\ and\ \bibinfo {author} {\bibfnamefont {C.}~\bibnamefont {Lee}},\
  }\href@noop {} {\bibfield  {journal} {\bibinfo  {journal} {Phys. Rev. B}\
  }\textbf {\bibinfo {volume} {55}},\ \bibinfo {pages} {10355} (\bibinfo {year}
  {1997})}\BibitemShut {NoStop}%
\bibitem [{\citenamefont {Perdew}\ \emph {et~al.}(1996)\citenamefont {Perdew},
  \citenamefont {Burke},\ and\ \citenamefont {Ernzerhof}}]{Perdew:1996}%
  \BibitemOpen
  \bibfield  {author} {\bibinfo {author} {\bibfnamefont {J.~P.}\ \bibnamefont
  {Perdew}}, \bibinfo {author} {\bibfnamefont {K.}~\bibnamefont {Burke}}, \
  and\ \bibinfo {author} {\bibfnamefont {M.}~\bibnamefont {Ernzerhof}},\
  }\href@noop {} {\bibfield  {journal} {\bibinfo  {journal} {Phys. Rev. Lett.}\
  }\textbf {\bibinfo {volume} {77}},\ \bibinfo {pages} {3865} (\bibinfo {year}
  {1996})}\BibitemShut {NoStop}%
\bibitem [{\citenamefont {Togo}\ \emph {et~al.}(2008)\citenamefont {Togo},
  \citenamefont {Oba},\ and\ \citenamefont {Tanaka}}]{phonopy}%
  \BibitemOpen
  \bibfield  {author} {\bibinfo {author} {\bibfnamefont {A.}~\bibnamefont
  {Togo}}, \bibinfo {author} {\bibfnamefont {F.}~\bibnamefont {Oba}}, \ and\
  \bibinfo {author} {\bibfnamefont {I.}~\bibnamefont {Tanaka}},\ }\href@noop {}
  {\bibfield  {journal} {\bibinfo  {journal} {Phys. Rev. B}\ }\textbf {\bibinfo
  {volume} {78}},\ \bibinfo {pages} {134106} (\bibinfo {year}
  {2008})}\BibitemShut {NoStop}%
\bibitem [{\citenamefont {B\"arnighausen}\ and\ \citenamefont
  {Handa}(1964)}]{Baernighausen1964}%
  \BibitemOpen
  \bibfield  {author} {\bibinfo {author} {\bibfnamefont {H.}~\bibnamefont
  {B\"arnighausen}}\ and\ \bibinfo {author} {\bibfnamefont {B.~K.}\
  \bibnamefont {Handa}},\ }\href@noop {} {\bibfield  {journal} {\bibinfo
  {journal} {J. Less-Comm. Metals}\ }\textbf {\bibinfo {volume} {6}},\ \bibinfo
  {pages} {226} (\bibinfo {year} {1964})}\BibitemShut {NoStop}%
\bibitem [{\citenamefont {Winter}\ \emph {et~al.}(2016)\citenamefont {Winter},
  \citenamefont {Li}, \citenamefont {Jeschke},\ and\ \citenamefont
  {Valent{\'{\i}}}}]{Winter2016}%
  \BibitemOpen
  \bibfield  {author} {\bibinfo {author} {\bibfnamefont {S.~M.}\ \bibnamefont
  {Winter}}, \bibinfo {author} {\bibfnamefont {Y.}~\bibnamefont {Li}}, \bibinfo
  {author} {\bibfnamefont {H.~O.}\ \bibnamefont {Jeschke}}, \ and\ \bibinfo
  {author} {\bibfnamefont {R.}~\bibnamefont {Valent{\'{\i}}}},\ }\href
  {\doibase 10.1103/physrevb.93.214431} {\bibfield  {journal} {\bibinfo
  {journal} {Phys. Rev. B}\ }\textbf {\bibinfo {volume} {93}},\ \bibinfo
  {pages} {214431} (\bibinfo {year} {2016})}\BibitemShut {NoStop}%
\bibitem [{\citenamefont {Winter}\ \emph {et~al.}()\citenamefont {Winter},
  \citenamefont {Riedl}, \citenamefont {Honecker},\ and\ \citenamefont
  {Valent{\'{\i}}}}]{Winter2017}%
  \BibitemOpen
  \bibfield  {author} {\bibinfo {author} {\bibfnamefont {S.~M.}\ \bibnamefont
  {Winter}}, \bibinfo {author} {\bibfnamefont {K.}~\bibnamefont {Riedl}},
  \bibinfo {author} {\bibfnamefont {A.}~\bibnamefont {Honecker}}, \ and\
  \bibinfo {author} {\bibfnamefont {R.}~\bibnamefont {Valent{\'{\i}}}},\
  }\href@noop {} {\bibinfo  {journal} {arXiv:1702.08466}\ }\BibitemShut
  {NoStop}%
\bibitem [{\citenamefont {Janssen}\ \emph {et~al.}(2016)\citenamefont
  {Janssen}, \citenamefont {Andrade},\ and\ \citenamefont
  {Vojta}}]{Janssen2016}%
  \BibitemOpen
\bibfield  {journal} {  }\bibfield  {author} {\bibinfo {author} {\bibfnamefont
  {L.}~\bibnamefont {Janssen}}, \bibinfo {author} {\bibfnamefont {E.~C.}\
  \bibnamefont {Andrade}}, \ and\ \bibinfo {author} {\bibfnamefont
  {M.}~\bibnamefont {Vojta}},\ }\href@noop {} {\bibfield  {journal} {\bibinfo
  {journal} {Phys. Rev. Lett.}\ }\textbf {\bibinfo {volume} {117}},\ \bibinfo
  {pages} {277202} (\bibinfo {year} {2016})}\BibitemShut {NoStop}%
\bibitem [{\citenamefont {Janssen}\ \emph {et~al.}()\citenamefont {Janssen},
  \citenamefont {Andrade},\ and\ \citenamefont {Vojta}}]{Janssen2017}%
  \BibitemOpen
  \bibfield  {author} {\bibinfo {author} {\bibfnamefont {L.}~\bibnamefont
  {Janssen}}, \bibinfo {author} {\bibfnamefont {E.~C.}\ \bibnamefont
  {Andrade}}, \ and\ \bibinfo {author} {\bibfnamefont {M.}~\bibnamefont
  {Vojta}},\ }\href@noop {} {\bibinfo  {journal} {unpublished}\ }\BibitemShut
  {NoStop}%
\bibitem [{\citenamefont {Majumder}\ \emph {et~al.}(2015)\citenamefont
  {Majumder}, \citenamefont {Schmidt}, \citenamefont {Rosner}, \citenamefont
  {Tsirlin}, \citenamefont {Yasuoka},\ and\ \citenamefont
  {Baenitz}}]{Majumder2015}%
  \BibitemOpen
\bibfield  {journal} {  }\bibfield  {author} {\bibinfo {author} {\bibfnamefont
  {M.}~\bibnamefont {Majumder}}, \bibinfo {author} {\bibfnamefont
  {M.}~\bibnamefont {Schmidt}}, \bibinfo {author} {\bibfnamefont
  {H.}~\bibnamefont {Rosner}}, \bibinfo {author} {\bibfnamefont {A.~A.}\
  \bibnamefont {Tsirlin}}, \bibinfo {author} {\bibfnamefont {H.}~\bibnamefont
  {Yasuoka}}, \ and\ \bibinfo {author} {\bibfnamefont {M.}~\bibnamefont
  {Baenitz}},\ }\href {\doibase 10.1103/physrevb.91.180401} {\bibfield
  {journal} {\bibinfo  {journal} {Phys. Rev. B}\ }\textbf {\bibinfo {volume}
  {91}},\ \bibinfo {pages} {180401} (\bibinfo {year} {2015})}\BibitemShut
  {NoStop}%
\end{thebibliography}%

\end{document}

\title{Supplemental Material: \\
       Field-induced quantum criticality in the Kitaev system $\alpha$-RuCl$_3$}

\author{A. U. B. Wolter}
\author{L. T. Corredor*}
\affiliation{Leibniz-Institut f\"ur Festk\"orper- und
Werkstoffforschung (IFW) Dresden, 01171 Dresden, Germany}
\thanks{These authors contributed equally to this work.}
\author{L. Janssen*}
\affiliation{Institut f\"ur Theoretische Physik, Technische
Universit\"at Dresden, 01062 Dresden,  Germany}
\author{K. Nenkov}
\affiliation{Leibniz-Institut f\"ur Festk\"orper- und
Werkstoffforschung (IFW) Dresden, 01171 Dresden, Germany}
\author{S. Sch\"{o}necker}
\affiliation{Department of Materials Science and Engineering, KTH
- Royal Institute of Technology, Stockholm 10044, Sweden}
\author{S.-H. Do}
\author{K.-Y. Choi}
\affiliation{Department of Physics, Chung-Ang University, Seoul
156-756, Republic of Korea}
\author{R. Albrecht}
\author{J. Hunger}
\author{T. Doert}
\affiliation{Fachrichtung Chemie und Lebensmittelchemie,
Technische Universit\"at Dresden, 01062 Dresden, Germany}
\author{M. Vojta}
\affiliation{Institut f\"ur Theoretische Physik, Technische
Universit\"at Dresden, 01062 Dresden,  Germany}
\author{B. B\"{u}chner}
\affiliation{Leibniz-Institut f\"ur Festk\"orper- und
Werkstoffforschung (IFW) Dresden, 01171 Dresden, Germany}
\affiliation{Institut f\"ur Festk\"orperphysik, Technische
Universit\"{a}t Dresden, 01062 Dresden, Germany}

\date{\today}

\maketitle

%%%%%%%%%%%%%%%%%%%%%%%%%%%%%%%%%%%%%%%%%%%%%%%%%%%%%%%%%%%%%%%%%%%%%%%%%%%%%%%%%%%

\section{Magnetic characterization of $\alpha$-R\lowercase{u}C\lowercase{l}$_3$}

\begin{figure}[!b]
\centering
\includegraphics[scale=0.46]{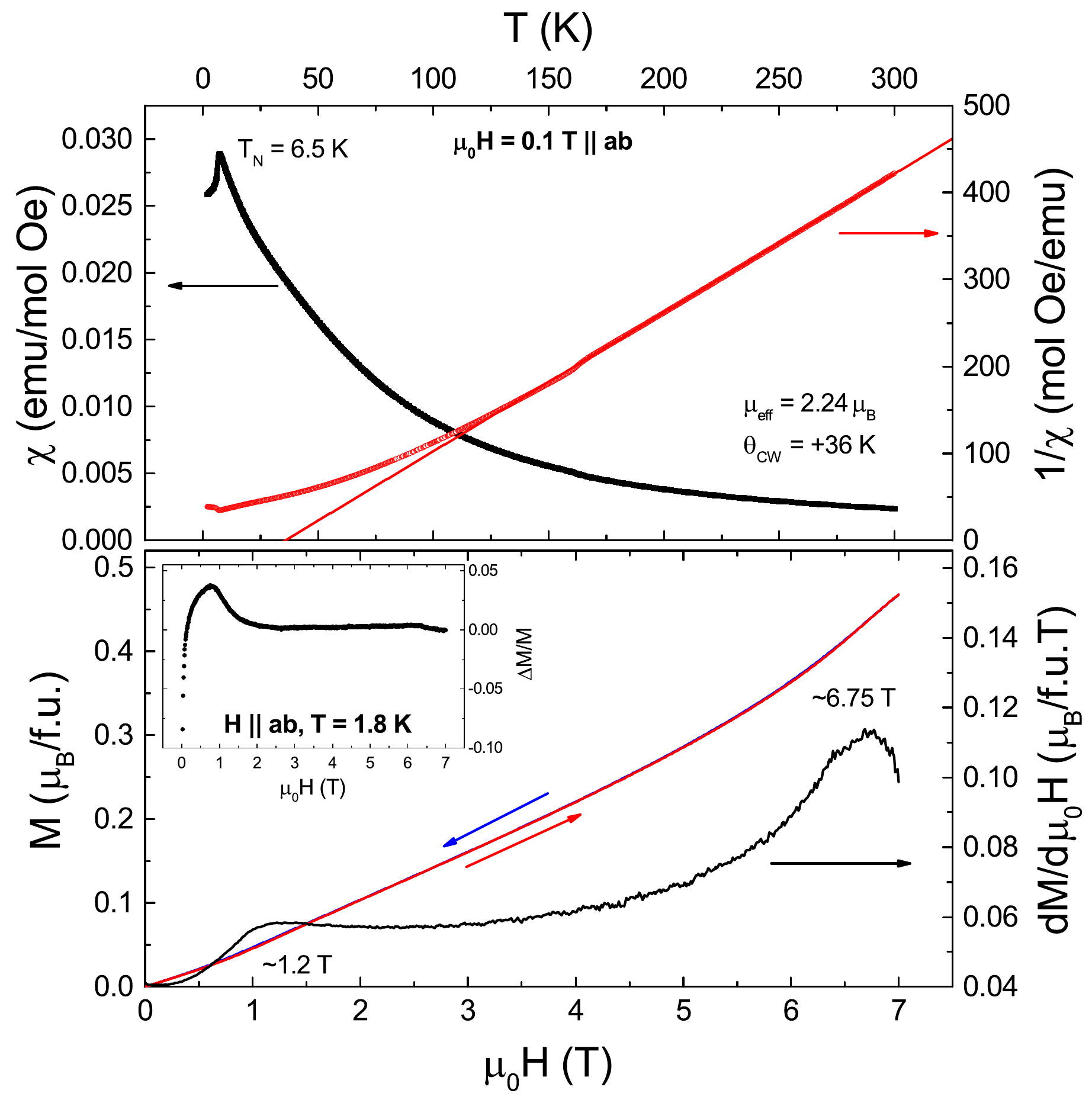}
\caption{(color online) Upper panel: The magnetic susceptibility
$\chi$ as function of temperature of \rucl for $\mu_0 H = 0.1$~T
$\parallel ab$ (left axis). On the right axis the inverse
susceptibility $1/\chi(T)$ is shown together with the Curie-Weiss
fit in the high-temperature regime. Lower panel: The magnetization
as function of field of \rucl measured at $1.8$~K (left axis)
together with its derivative $dM/d(\mu_0 H)$ (right axis). In the
inset the relative difference of the magnetization for up-and
down-sweeps of the magnetic field $\Delta M/M$ is depicted as
function of field. } \label{fig:magnetization}
\end{figure}

The temperature dependence of the magnetic susceptibility
$\chi(T)$ of \rucl is shown in Fig.~\ref{fig:magnetization} (upper
panel) for $\mu_0 H$ = 0.1 T $\parallel$ {\it{ab}}. Note that the
same single crystal was used for the magnetic characterization and
the specific heat capacity measurements. Clearly, $\chi(T)$
exhibits a sharp maximum at $\approx 7.2$~K in agreement with
earlier reports on high-quality single crystals, which only have a
very small amount of stacking faults
~\cite{Banerjee2016,Cao2016,Baek2017}. From the derivative
$d(\chi\cdot T)/dT$ the transition temperature signalling the
transition into the magnetically long-range ordered state is
determined to $T_N = 6.5$~K.

From the temperature dependence of the inverse susceptibility (red
line in the upper panel of Fig.~\ref{fig:magnetization}) a linear
scaling of 1/$\chi(T)$ with temperature $T$ is observed for $T>T_s
\approx 160$~K. $T_s$ marks the first-order structural transition
of \rucl \cite{Baek2017}. From a fit of the inverse susceptibility
to a Curie-Weiss law, a Curie-Weiss temperature $\theta_{\rm CW} =
+36$~K and an effective magnetic moment $\mu_{\rm eff} = 2.24
\mu_B$ were extracted for $H \parallel ab$. Notably, the effective
moment is much larger than the spin-only value of 1.73$\mu_B$
expected for Ru$^{3+}$, pointing towards a large orbital
contribution to the magnetic moment.

The magnetization of \rucl as function of field $H \parallel ab$
measured at $1.8$~K is depicted in the lower panel of Fig.~
\ref{fig:magnetization}. From the derivative curve $dM/d(\mu_0 H)$
two changes of slope can clearly be discerned at $\mu_0 H \approx
1.2$~T and $\mu_0 H \approx 6.75$~T. While the transition around
$6.75$~T is in line with the field-induced QCP observed in our
specific-heat study in this work, the one around 1.2~T is still a
matter of debate. Following the change of slope of $M(H)$ in the
low-field regime together with the magnetic susceptibility at
lowest $T$, the presence of paramagnetic impurities can be
discarded as origin for the low-field anomaly around $1.2$~T.
Rather, the anomaly could be due to a redistribution in domain
population occurring in this rather low field
range~\cite{Sears2017}.

Looking at the hysteretic behavior of our magnetization curves for
up- and down-sweeps of the magnetic fields, no substantial
hysteresis can be observed for fields above $\sim 2$~T. This is in
perfect agreement with our field-induced QCP scenario at $\mu_0
\Hc$ $\approx$ 6.9~T, and underlines the second-order nature of
the phase transition at $\mu_0 \Hc$.

%%%%%%%%%%%%%%%%%%%%%%%%%%%%%%%%%%%%%%%%%%%%%%%%%%%%%%%%%%%%%%%%%%%%%%%%%%%%%%%%%%%

\section{Phonon calculations for R\lowercase{h}C\lowercase{l}$_3$}

\subsection{Computational details}

The first-principles calculations were performed with the
projector-augmented wave method as implemented in the Vienna
\emph{ab initio} simulation package
(VASP)~\cite{Bloechl:1994,Kresse:1999,Kresse:1996}. The
force-constant matrix was obtained through the super cell approach
within the finite displacement
method~\cite{Parlinski:1997,Chaput:2011} taking into account
non-analytical term corrections~\cite{Gonze:1997b}. The
generalized-gradient approximation in the parameterization of
Perdew, Burke, and Ernzerhof (PBE)~\cite{Perdew:1996} was adopted
to describe exchange and correlation. The software PHONOPY was
employed to determine the phonon dispersion relations and the
phonon density of states (DOS) from the force-constant matrix, as
well as the heat capacity at constant volume~\cite{phonopy}. The
experimental single crystal structure parameters for RhCl$_3$ were
used in the calculations, which confirm the literature
data~\cite{Baernighausen1964}.

The convergence of all numerical parameters was carefully checked.
All VASP calculations were carried out with the global precision
switch ``Accurate'' employing a plane-wave cutoff of $400$\,eV.
The grid for augmentation charges contained eight times the
default number and the convergence criteria for the total energy
was set to $10^{-8}$\,eV. $\Gamma$-point calculations for a
$4\times4\times4$ super cell in terms of the conventional eight
atoms unit cell (corresponding to a $4\times4\times4$ phonon grid
partitioning) mesh were adopted for the present results.

\section{Results}

\begin{figure}
\resizebox{\columnwidth}{!}{\includegraphics[clip]{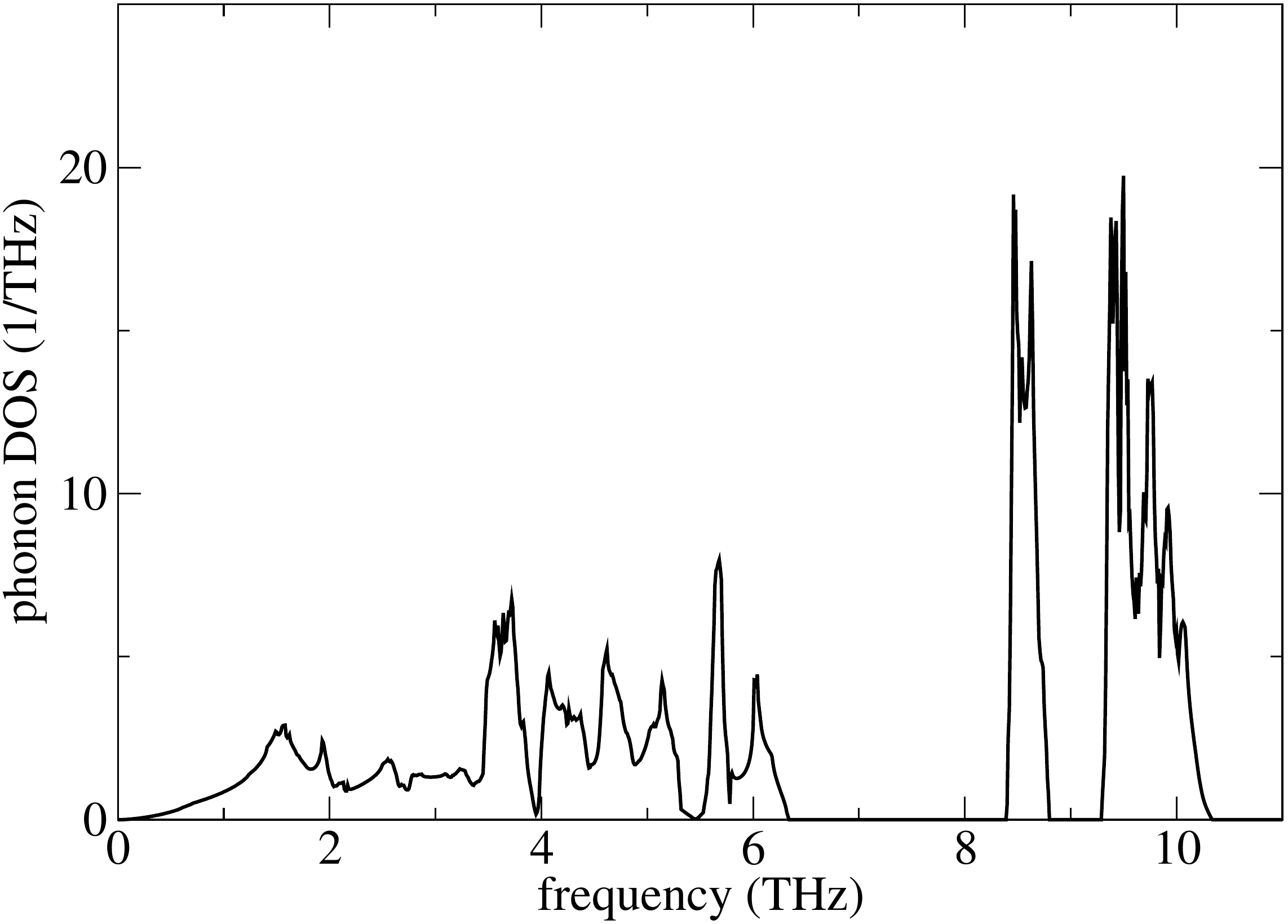}}
\caption{\label{fig:pDOS}Phonon DOS for RhCl$_{3}$. The DOS is
normalized to the number of normal modes per primitive unit cell.}
\end{figure}

\begin{figure}
\resizebox{\columnwidth}{!}{\includegraphics[clip]{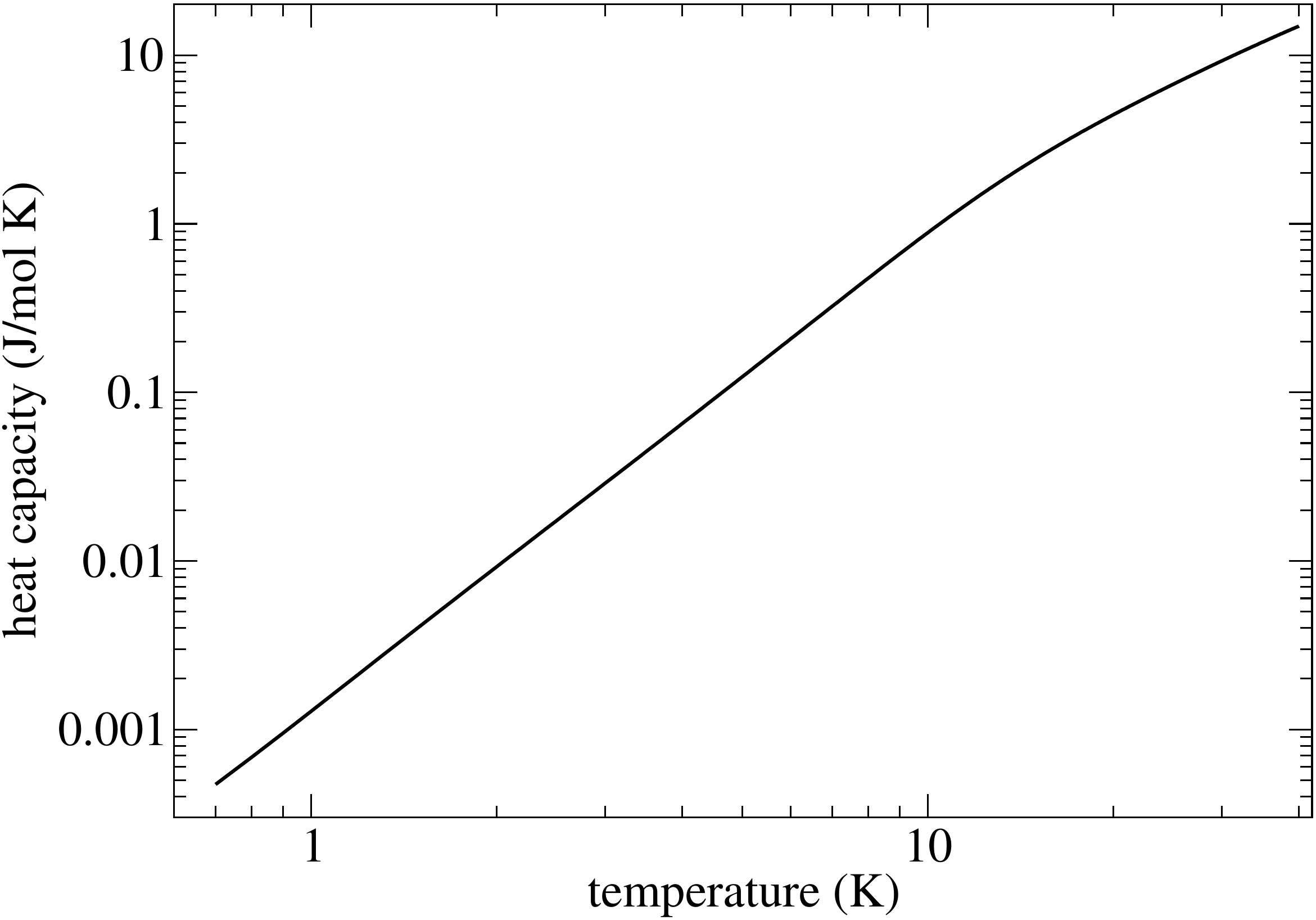}}
\caption{\label{fig:Cv}Log-log plot of the heat capacity at
constant volume for RhCl$_{3}$. The heat capacity is given per
formula unit.}
\end{figure}

The computed phonon DOS and the derived heat capacity in the low
temperature region for RhCl$_{3}$ are shown in
Figs.~\ref{fig:pDOS} and~\ref{fig:Cv}, respectively. As is
evident, the phonon spectrum is gapped twice, exhibits a
Debye-like low-frequency behavior, and possesses a band width of
approximately 10.3\,Thz. The temperature dependence of the heat
capacity follows a Debye-like $T^3$ behavior up to approximately
10\,K.

%%%%%%%%%%%%%%%%%%%%%%%%%%%%%%%%%%%%%%%%%%%%%%%%%%%%%%%%%%%%%%%%%%%%%%%%%%%%%%%%%%%

\section{Field-induced QCP in {\JKG} honeycomb lattice model}

\subsection{Modelling}

To date, the debate about the most appropriate effective spin
model to describe the magnetic behavior of \rucl has not been
settled. Most proposals involve nearest-neighbor Heisenberg,
Kitaev, and symmetric off-diagonal exchanges on a two-dimensional
honeycomb lattice; often second- and/or third-neighbor
interactions are invoked as well.
Below we will show results for a concrete minimal model derived
from ab-initio density functional theory, containing
nearest-neighbor Heisenberg $J_1$, Kitaev $K_1$, and off-diagonal
$\Gamma_1$ interaction as well as a third-nearest-neighbor
Heisenberg $J_3$ interaction \cite{Winter2016}:
%\begin{widetext}
\begin{align}
 \mathcal H &=
 \sum_\text{1st nn} \left[
 J_1 \vec S_i \cdot \vec S_j
 + K_1 S_i^\gamma S_j^\gamma
 + \Gamma_1 (S_i^\alpha S_j^\beta + S_i^\beta S_j^\alpha)\right] \notag \\
 & +
 \sum_\text{3rd nn} J_3 \vec S_i \cdot \vec S_j.
\label{eq:jkg}
\end{align}
%\end{widetext}
%
Here, $\{\alpha, \beta, \gamma\} = \{x, y, z\}$ on a
nearest-neighbour $z$ bond, for example. The spin quantization
axes point along the cubic axes of the RuCl$_6$ octahedra, such
that the $[111]$ direction is perpendicular to the honeycomb $ab$
plane (sometimes referred to as $c^*$ axis) and the in-plane
$[\bar 1 1 0]$ direction points along a Ru-Ru nearest-neighbor
bond of the honeycomb lattice. Trigonal distortion is neglected in
this simple model.
The values for the exchange couplings can be estimated from the
\textit{ab initio} calculations~\cite{Winter2016}; however, we
find better agreement with our experimental data by using a
slightly adapted parameter set that has recently been suggested by
comparing with neutron scattering data (at zero
field)~\cite{Winter2017}:
\begin{align} \label{eq:couplings}
 (J_1, K_1, \Gamma_1, J_3) & = \left(-0.5, -5.0, +2.5, +0.5\right)\,\text{meV}.
\end{align}

We are interested in the behavior of this model in the presence of
an external magnetic field, i.e., described by the Hamiltonian
$\mathcal H' = \mathcal H - g \mu_0 \mu_\mathrm{B} \sum_i \vec H
\cdot \vec S_i$. Here, $g \mu_\mathrm{B} \vec S$ corresponds to
the effective moment of the $J_\text{eff} = 1/2$ states in the
crystal. Solving this (or other relevant) models for
quantum-mechanical spins $1/2$ requires large-scale numerics, and
detailed studies in an applied field are lacking.

\subsection{Spin-wave theory for $H>\Hc$}

The model \eqref{eq:jkg} can be solved in the semiclassical limit
of large spin $S$ \cite{Janssen2016, Janssen2017}. At zero field,
it has a zigzag antiferromagnetic ground state. At finite $\vec H
\parallel [\bar 1 1 0] \in ab$, the zigzag state cants towards the
magnetic field. At a critical field strength $\Hc$, there is a
continuous transition towards a (partially) polarized high-field
phase. For the critical field we find, in the semiclassical limit,
$\mu_0 \Hc = 0.586 \frac{|K_1 S|}{g \mu_\mathrm{B}} \simeq
9\,\mathrm{T}$ if we assume the previously estimated $g$ factor of
$g \simeq 2.8$~\cite{Majumder2015}.
Given the fact that our model does not include any free fitting
parameter and in light of the semiclassical approximation we find
the rough agreement with our experimental finding of $\mu_0 \Hc
\simeq 6.9\,\mathrm{T}$ satisfactory.

\begin{figure*}
 \includegraphics[scale=.75]{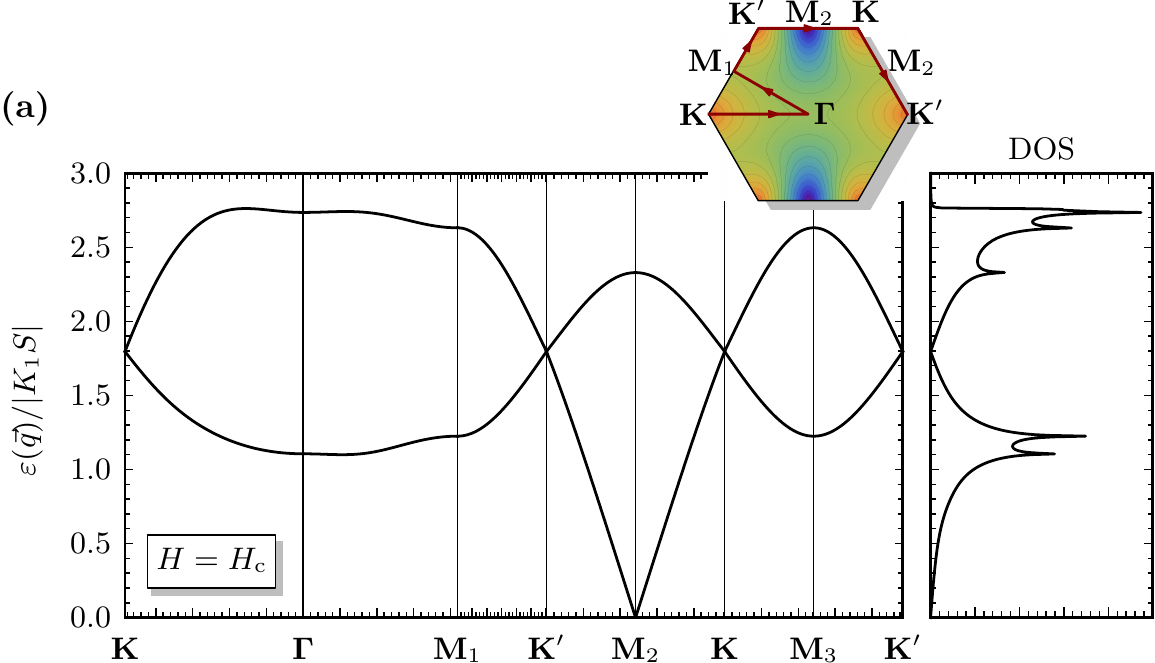}\hfill
 \includegraphics[scale=.75]{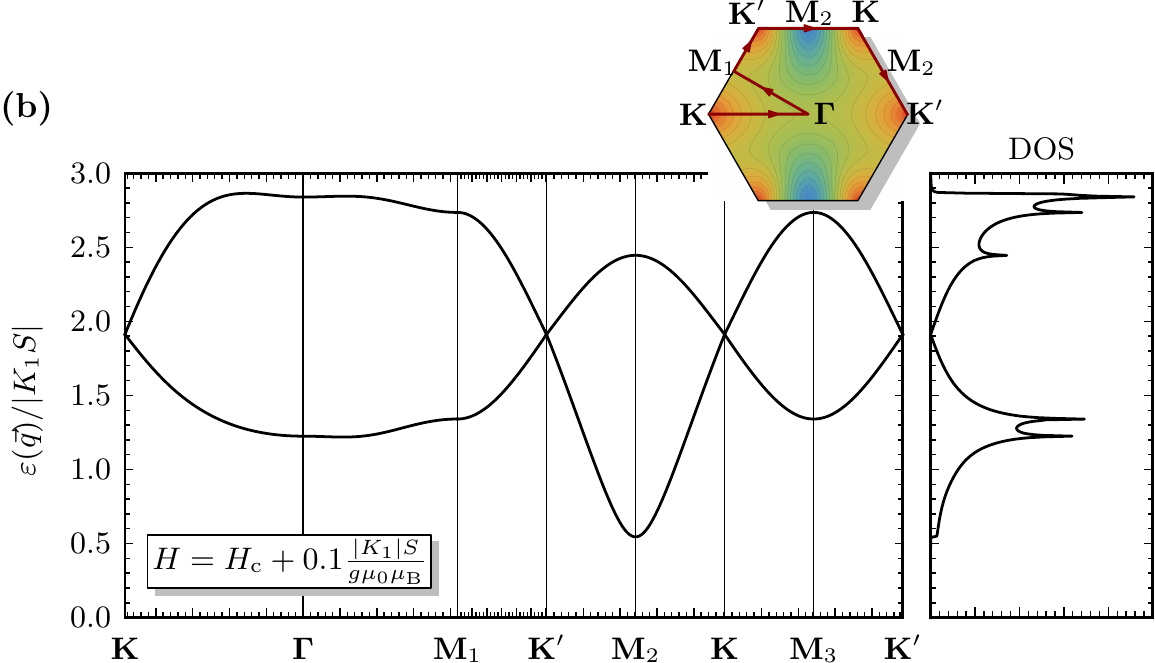}

 \caption{(color online) Calculated excitation spectra and magnon DOS at the quantum critical point (a) and in high-field phase (b), respectively, for $J_1 - K_1 - \Gamma_1 - J_3$ model on honeycomb lattice in external field $\vec H \parallel [\bar 1 1 0]$. The inset shows the lower-band dispersion in the first Brillouin zone (color plot) and the path along the high-symmetry lines used in the main panel (red line).
 We have used $(J_1, K_1, \Gamma_1, J_3) = (-0.5, -5.0, +2.5, +0.5)\,\text{meV}$.
 }
 \label{fig:spectrum}
\end{figure*}

\begin{widetext}
The excitation spectrum in the high-field phase can be computed
within spin-wave theory. We employ the Holstein-Primakoff
representation
\begin{align}
 \vec S_i & =
 \begin{cases}
 (S - a^\dagger_i a_i) \vec n
 + \sqrt{\frac{S}{2}} (a_i + a_i^\dagger) \vec e
 + \iu \sqrt{\frac{S}{2}} (a_i - a_i^\dagger) (\vec n \times \vec e) + \mathcal O(1/\sqrt{S}), &
 \qquad \text{if } i \in \mathrm{A}, \\
 (S - b^\dagger_i b_i) \vec n
 + \sqrt{\frac{S}{2}} (b_i + b_i^\dagger) \vec e
 + \iu \sqrt{\frac{S}{2}} (b_i - b_i^\dagger) (\vec n \times \vec e) + \mathcal O(1/\sqrt{S}), &
 \qquad \text{if } i \in \mathrm{B},
 \end{cases}
\end{align}
with $\vec n = (-\vec e_x+\vec e_y)/\sqrt{2} \parallel \vec H$ and
$\vec e = -\vec e_z$. $\vec e_x$, $\vec e_y$, and $\vec e_z$ are
the spin quantization axes. $a_i^\dagger$ and $a_i$ ($b_i^\dagger$
and $b_i$) are the magnon creation and annihilation operators at
site $i$ on sublattice A (B). To the leading order in $1/S$, we
find the spin-wave Hamiltonian
\begin{align}
 \mathcal H_\text{SW} & = S \sum_{\vec q \in \mathrm{BZ}}
 \left[
 \epsilon_0 \left(a^\dagger_{\vec q} a_{\vec q} + b^\dagger_{\vec q} b_{\vec q}\right)
 + \lambda_0(\vec q) a^\dagger_{\vec q} b_{\vec q} + \lambda_0^*(\vec q) b^\dagger_{\vec q} a_{\vec q}
 + \lambda_1(\vec q) a_{-\vec q} b_{\vec q} + \lambda_1^*(-\vec q) a^\dagger_{\vec q} b^\dagger_{- \vec q}
 \right],
\end{align}
with the coefficients
\begin{align}
 \epsilon_0 & = g \mu_0 \mu_\mathrm{B} H/S - 3 J_1 - K_1 + \Gamma_1, \\
 \lambda_0(\vec q) & =
 \left(J_1+ \frac{K_1}{4}\right) \left(\ee^{\iu \vec q \cdot \vec \delta_x} + \ee^{\iu \vec q \cdot \vec \delta_y}\right)
 + \left(J_1+ \frac{K_1}{2} + \frac{\Gamma_1}{2}\right) \ee^{\iu \vec q \cdot \vec \delta_z} +
 J_3 \left(\ee^{-2\iu \vec q \cdot \vec \delta_x} + \ee^{-2\iu \vec q \cdot \vec \delta_y} + \ee^{-2\iu \vec q \cdot \vec \delta_z}\right), \\
 \lambda_1(\vec q) & =
 \left(-\frac{K_1}{4} + \frac{\iu \Gamma_1}{\sqrt{2}}\right) \left(\ee^{\iu \vec q \cdot \vec \delta_x} + \ee^{\iu \vec q \cdot \vec \delta_y}\right)
 + \frac{K_1 - \Gamma_1}{2} \ee^{\iu \vec q \cdot \vec \delta_z}.
\end{align}
$\mathcal H_\mathrm{SW}$ can be diagonalized by means of a
Bogoliubov transformation. The resulting excitation spectrum
together with the corresponding density of states (DOS) for the
parameter set of Eq.~\eqref{eq:couplings} is displayed for two
different values of the magnetic field at and above the quantum
critical point (QCP) in Fig.~\ref{fig:spectrum}. The spectrum is
gapped for any $H > \Hc =  0.586 \frac{|K_1 S|}{g \mu_0
\mu_\mathrm{B}}$ (in agreement with the classical critical field
strength) with a gap value of
\begin{align} \label{eq:gap}
 \Delta(H) = 1.30 \left\lvert K_1 S \right\rvert \left(\frac{H-\Hc}{\Hc}\right)^{1/2} + \mathcal O\!\left[\left((H-\Hc)/\Hc\right)^{3/2}\right],
\end{align}
which is roughly of the order of magnitude of the experimentally
observed gap. As quantum effects are enhanced at low energies, we
expect Eq.~\eqref{eq:gap} to receive sizable corrections when
magnon interactions are taken into account. In particular, the
true gap exponent $\nu z$ will deviate from the mean-field value
$(\nu z)_\text{MF} = 1/2$ we have obtained here. This prevents a
more detailed quantitative comparison with the experimental gap
behavior.
\end{widetext}

\begin{figure*}
 \includegraphics[scale=1]{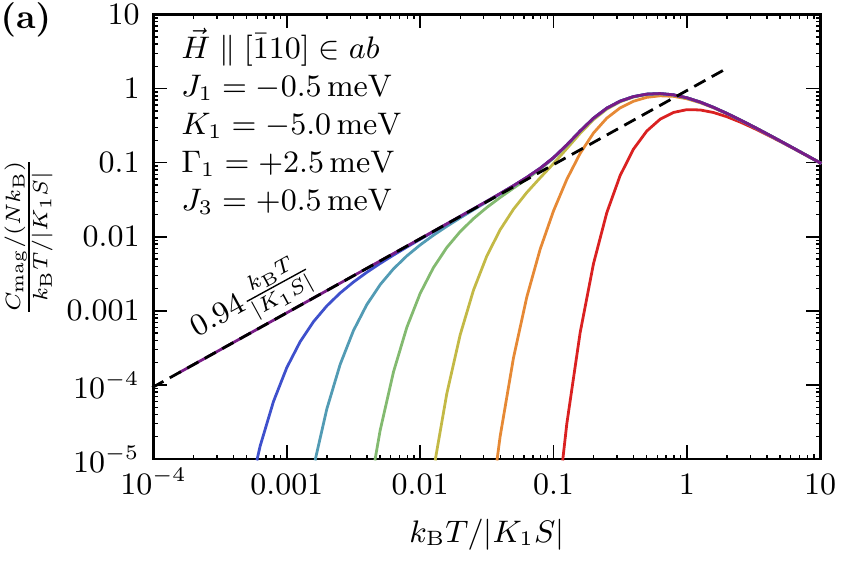} \hfill
 \includegraphics[scale=1]{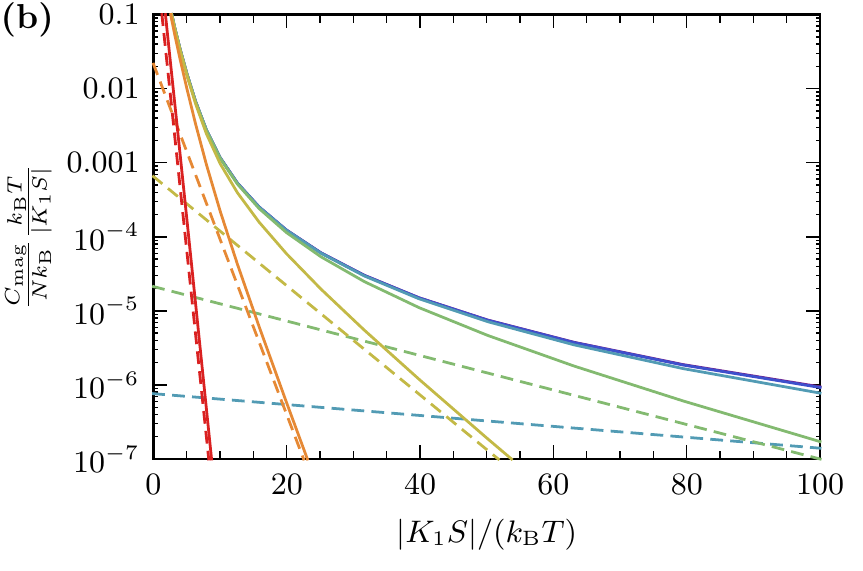} \\
 \raggedright \includegraphics[scale=1]{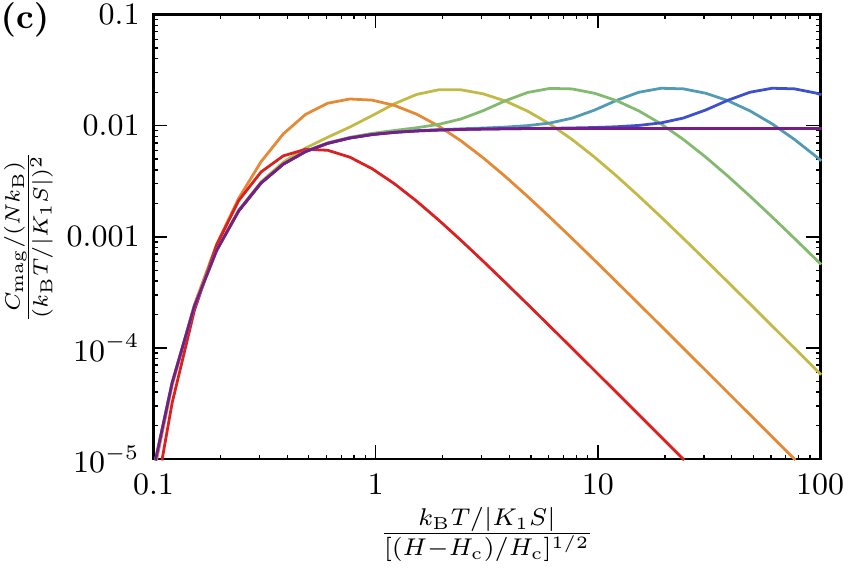}
 \includegraphics[scale=1]{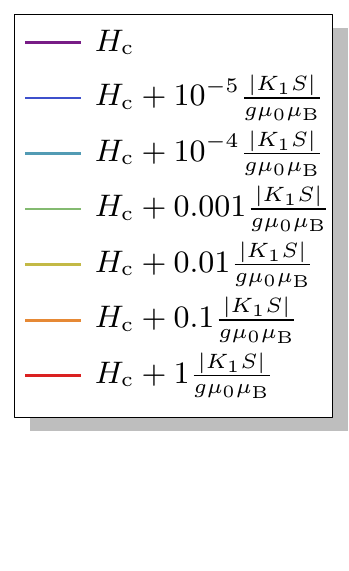}\hfill

 \caption{(color online)
 (a) Double-log plot of the specific heat $\Cmag/T$ versus temperature $T$ for a honeycomb-lattice {\JKG} model in external field $\vec H \parallel [\bar 1 1 0]$ for different magnetic field strengths $H \geq \Hc$. At the quantum critical point $H = \Hc$ and low $T$, the specific heat scales as $C_\text{mag} \propto T^{d/z}$ with dimensionality $d=2$ and the dynamical critical exponent $z = 1$ (dashed line).
 (b) Same data as (a), but now plotted as $\Cmag T$ versus $1/T$ in log-linear plot. The dashed lines show the low-$T$ approximation according to Eq.~\eqref{eq:cXT}.
 (c) Scaling plot $\Cmag/T^{d/z}$ versus $T/(H - \Hc)^{\nu z}$ with correlation-length exponent $\nu$. For our model, we have $\nu=1/2$ at the level of the present mean-field-like approximation.}
 \label{fig:specific-heat}
\end{figure*}

We note, however, that thermodynamic quantities, such as the
specific heat at low to intermediate temperatures, should be
expected to be lesser affected by our linear spin-wave
approximation, since they predominantly depend on the parts of the
excitation spectrum with a large density of states, and these are
located at higher energy.

\subsection{Specific heat for $H>\Hc$}

\begin{figure}
\includegraphics[scale=.9]{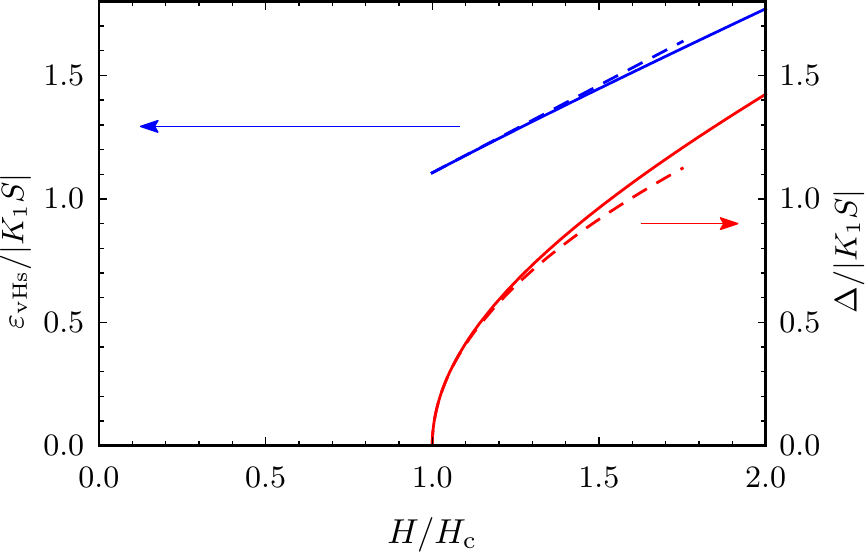}
\caption{(color online) Red: Calculated gap $\Delta(H)$ as
function of magnetic field $H/\Hc \geq 1$. Blue: Energy of the
first van-Hove singularity $\varepsilon_\mathrm{vHs}$. The dashed
curves correspond    to expansions in small $(H-\Hc)/\Hc$,
yielding $\Delta(H)/|K_1 S| \simeq 1.30 [(H-\Hc)/\Hc]^{1/2}$
(Eq.~\eqref{eq:gap}) and $\varepsilon_\mathrm{vHs}/|K_1 S| \simeq
1.11 + 0.71 (H-\Hc)/\Hc$, respectively. } \label{fig:vhs}
\end{figure}

The heat capacity is obtained from the spectrum via
\begin{align}
\Cmag(T,H)  & = \sum_{\alpha = 1,2} \sum_{\vec q \in \mathrm{BZ}}
 \frac{\partial}{\partial T} \frac{\varepsilon_{\alpha}(\vec q)}{\exp\left[\varepsilon_{\alpha}(\vec q)/(k_\mathrm{B} T)\right] - 1},
\end{align}
where $\varepsilon_{1,2}(\vec q)$ are the two magnon bands. The
result is given for different magnetic field strengths in
Fig.~\ref{fig:specific-heat}(a).
At low temperatures, and $H$ not too close to $\Hc$, the specific
heat is exponentially suppressed,
\begin{align} \label{eq:cXT}
\Cmag(T,H) \simeq k_\mathrm{B} \left(\frac{\rho_0
\Delta^2}{k_\mathrm{B} T}\right) \ee^{-\Delta/(k_\mathrm{B} T)},
\quad
 \text{for } k_\mathrm{B} T \ll \Delta(H),
\end{align}
where $\rho_0 \equiv \rho_0(H)$ is the density of states at the
band minimum. This is shown in Fig.~\ref{fig:specific-heat}(b).
Close to the QCP, on the other hand, the critical part of the
specific heat is expected to follow a scaling law
\begin{align}
\Cmag(T,H) = T^{d/z} f_\pm \!\left(T/(H-\Hc)^{\nu z}\right)
\end{align}
with the spatial dimensionality $d=2$, the dynamical critical
exponent $z=1$, the correlation-length exponent $\nu$, and scaling
functions $f_\pm(x)$ above $(+)$ and below $(-)$ the QCP. This is
demonstrated for our theoretical data in
Fig.~\ref{fig:specific-heat}(c).
As a consequence, directly at the QCP for $H = \Hc$, the specific
heat follows a power law at low temperatures, $\Cmag(T,H) \propto
T^{2}$, see dashed line in Fig.~\ref{fig:specific-heat}(a). For
fields $H>\Hc$ the low-$T$ specific heat is gapped, with a gap
which depends sublinearly on $(H-\Hc)$, see Fig.~\ref{fig:vhs}.

Interestingly, $\Cmag/T$ displays a maximum at higher
temperatures, $k_\mathrm{B} T \sim \mathcal O(|K_1 S|)$. The
position of this maximum shifts approximately linearly with $H$;
this can be attributed to the shift of the high-energy part of the
spectrum that has a large weight, such as the location of the
van-Hove singularities at $\varepsilon_\text{vHs} \sim \mathcal
O(|K_1 S|)$ at $H=\Hc$. The shift of $\varepsilon_\text{vHs}$ with
field is illustrated in Fig.~\ref{fig:vhs}. Note that the weight
near $\varepsilon_\text{vHs}$ is particularly large due to almost
flat portions of the magnon bands, arising from the combination of
$K_1$ and $\Gamma_1$ terms.

We emphasize that it is this specific-heat maximum which limits
the validity of scaling in our theoretical data,
Fig.~\ref{fig:specific-heat}(c). This is not unlike what happens
in the experimental data where scaling is spoiled by the presence
of a small energy scale in the magnon spectrum. Spectroscopic
investigations of the excitation spectrum at elevated fields are
clearly called for.

%%%%%%%%%%%%%%%%%%%%%%%%%%%%%%%%%%%%%%%%%%%%%%%%%%%%%%%%%%%%%%%%%%%%%%%%%%%%%%%%%%%

\bibliography{aRuCl3_Paper_v1_LT}